\documentclass{aa}  

\usepackage{graphicx}
\usepackage{txfonts}
\usepackage{lscape}
\usepackage{float}
\usepackage{hyperref}
\begin{document} 

   \title{Multiple star-forming episodes of intermediate-redshift galaxies}

   \author{C. Mu\~{n}oz L\'{o}pez
          \inst{1}\fnmsep\thanks{e-mail: \texttt{cmunoz@aip.de}}
          \and
          Davor Krajnovi\'{c}
          \inst{1}
          \and
          B. Epinat
          \inst{2,3}
          \and
          T. Urrutia
          \inst{1}
          \and
          I. Pessa
          \inst{1}
          \and
          T. Contini
          \inst{4}
          \and
          T. Nanayakkara
          \inst{5}
          \and 
          J. Pharo
          \inst{1}
          \and
          T. S. Gon\c{c}alves
          \inst{6}
          \and
          Tran Thi Thai
          \inst{7}
          \and
          N. F. Bouch\'{e}
          \inst{8}
          .
          }

   \institute{Leibniz-Institut f\"ur Astrophysik Potsdam (AIP),
        An der Sternwarte 16, 14482 Potsdam, Germany
        \and
        Canada-France-Hawaii Telescope, 65-1238 Mamalahoa Highway, Kamuela, HI 96743, USA
        \and
        Aix Marseille Univ, CNRS, CNES, LAM, Marseille, France
        \and
        Institut de Recherche en Astrophysique et Planétologie (IRAP), Université de Toulouse, CNRS, UPS, CNES, Toulouse, France
        \and
        Centre for Astrophysics and Supercomputing, Swinburne University of Technology, P.O. Box 218, Hawthorn, 3122, VIC, Australia
        \and
        Universidade Federal do Rio de Janeiro, Observatório do Valongo, Ladeira do Pedro Ant\^{o}nio, 43, Saúde CEP 20080-090 Rio de Janeiro, RJ, Brazil
        \and
        National Astronomical Observatory of Japan, 2-21-1 Osawa, Mitaka, Tokyo 181-8588, Japan
        \and
        Univ. Lyon1, ENS de Lyon, CNRS, Centre de Recherche Astro-physique de Lyon (CRAL) UMR5574, F-69230 Saint-Genis-Laval, France
        }

   \date{Received 17 April 2025 / Accepted 18 August 2025}
 
  \abstract
  {In this work, we derive and analyse the star formation histories of 393 intermediate-redshift (0.1 $\leq$ z $\leq$ 0.9) galaxies with stellar masses between $\sim$10$^{8}$ - 10$^{12}$ M$_{\odot}$. We focus on galaxies  located in the CANDELS/GOOD-S and CANDELS/COSMOS fields that have been observed with different surveys using MUSE. We  probe a cosmic time of approximately 6 billion years (Gyr) and a range of environments, from field (low-density systems) to rich groups (high-density systems). We find that the galaxies' stellar mean ages, metallicities, and star formation rates (SFRs) follow similar trends to galaxies as those characterising the nearby Universe. We modelled the derived SFRs, quantifying and characterising the number of star-forming episodes (SFEs). We found that more than 85$\%$ of the galaxies have more than one event of star formation, typically described with an exponentially decaying SFR and subsequent Gaussian-like episode(s) of star formation. We also observe that massive galaxies have fewer SFEs than low-mass systems and that they form their stellar mass and reach quiescence faster than lower mass galaxies. Moreover, the history of mass assembly for the most massive galaxies in the sample can be described with only one episode of star formation in the early Universe, which we detected as an exponential decrease that was longer in duration than subsequent SF events. This early event has typically been completed by z $\sim$ 3 and it accounts for a high fraction of the total stellar mass, from $\sim$40$\%$ for low-mass galaxies to more than 50$\%$ for higher-mass galaxies. We also analysed the dependence of stellar population parameters with the various environments probed by the sample, finding no significant correlations between different group environments; however, our field galaxies are generally distinct from group galaxies in terms of the mass, metallicity, stellar ages, and formation timescales. We discuss possible biases in the sample selection and examine how representative our galaxies are of the overall galaxy population at the targeted redshifts.}

   \titlerunning{Multiple star-forming episodes and star formation histories at 0.1 < z < 0.9 in MUSE surveys}
   \authorrunning{Constanza Mu\~{n}oz L\'{o}pez et al.}
   \maketitle
%
\section{Introduction}

Studying the stellar population of galaxies offers insights into how the history of mass assembly shaped their observed state. Observations enable us to derive the star formation history (SFH) and chemical composition of these systems. The first works studying the stellar population in galaxies were obtained through the analysis of photometric data around 1960 \citep{1968adga.book.....S}. Through the study of galaxy's optical colours \citep{1959PASP...71..106B, 1981A&A...100L..20V, 2001AJ....122.1861S, 2003ApJS..149..289B}, using colour-magnitude relations, galaxies can be classified into three main categories: blue cloud, which are galaxies where their luminosity is dominated by bluer wavelengths; red sequence, a narrow band of systems with red colours; and in between the green valley \citep{2003MNRAS.341...33K}. Studying the colours of their current stellar population allows us to infer the SFH of galaxies \citep{1968ApJ...151..547T, 1972A&A....20..383T, 1973ApJ...179..427S, 1986A&A...161...89S}. However, galaxy colours are strongly affected by degeneracies. The age-metallicity degeneracy \citep{1999ASPC..192..283W} does not permit disentangling if the colour of a galaxy is due to the ages or the metallicity of its stellar population. Moreover, ultraviolet (UV), and infrared (IR) light are sensitive to different ranges of star formation timescales, which depend on the wavelength and SFH, known as SFH-dust degeneracy \citep{2001ApJ...559..620P}. 

Different absorption lines (e.g. Ca H+K, Balmer lines, Mg b, Fe, etc.) are useful for studying stellar populations with various ages and metallicities and star formation of galaxies \citep{2014ApJ...792...95C, 2014ApJ...788...72G}. Nevertheless, most of the knowledge on the SFH and chemical compositions of galaxies can be retrieved from spectroscopic data using absorption features. Initially, large spectroscopic studies of stellar populations in the Local Universe were based on line indices of specific absorption features \citep{1992ApJ...398...69W, 1995IAUS..164..249F, 2000AJ....119.1645T}, while their interpretation was based on stellar population synthesis (SPS) models. Subsequently, with improvements in SPS and computational speed, studies shifted towards a full-spectrum fitting approach \citep{2003MNRAS.343.1145P, 2005MNRAS.358..363C, 2011MNRAS.417.1643K, 2015MNRAS.448.3484M, 2016A&A...592A..19C, 2017A&A...607A.128G} or a hybrid method of the full-spectrum fitting within an index (e.g.\citealp{2010MNRAS.404.1775T}, \citealp{2019MNRAS.484.3425M}), as well as a combination of spectra and photometric bands \citep{2023MNRAS.526.3273C}.

Several surveys of nearby galaxies (e.g. ATLAS$^{3D}$, \citealp{2011MNRAS.413..813C}; MaNGA, \citealp{2015ApJ...798....7B}; SAMI, \citealp{2012MNRAS.421..872C}; CALIFA, \citealp{2012A&A...538A...8S}) have studied the stellar population and SFHs of galaxies in the nearby Universe using spectrocopic data. \cite{2015MNRAS.448.3484M} analysed the stellar population of early-type galaxies from the ATLAS$^{3D}$ survey using line-strength indices and single stellar population models. They found that at fixed-mass, compact systems have on average older, more metal-rich, and higher alpha-element enhanced stellar populations compared to their larger counterparts. Also, they found that the duration of star formation (SF) is systematically longer in lower-mass galaxies. They observed that the stellar population properties vary only with the galaxy mass and are independent of the environment. \citet{2010ApJ...721..193P}, using a sample of galaxies from SDSS, zCOSMOS, and other deep surveys, found that the differential effects of environment and mass are separable up to z $\sim$ 1 in the quenching processes of galaxies. Among these two parameters, the quenching can primarily be driven by mass, but the environment also plays a role in the process of large-scale structure developing in the Universe. 

Similar results to those published by \cite{2015MNRAS.448.3484M} regarding the scatter in the age-M$_{\star}$ and metallicity-M$_{\star}$ relations were reported in \citet{2017MNRAS.472.2833S} and \citet{2018MNRAS.476.1765L} using SAMI and MaNGA data including galaxies of all morphologies, respectively. These surveys mainly target  galaxies with stellar mass M$_{\star}$ $\geq$ 10$^{9}$M$_{\odot}$. In lower mass galaxies (M$_{\star}$ $\leq$ 10$^{8}$M$_{\odot}$), used multi-band SEDs \cite{2012AJ....143...47Z} to derive the star formation rate (SFR) over a more recent timescale (i.e. 0.1 Gyr and 1 Gyr) along with  the stellar mass surface density profiles. They found that nearby dwarf irregular (dIrr) galaxies have much more complex SFHs than larger spiral galaxies.

Understanding galaxy evolution and its dependence on the evolution of the Universe calls for studies at higher redshifts. Using DEEP2 survey data, \citet{2015MNRAS.454.1332S} studied the SFHs of 154 galaxies in a redshift range 0.7 $\leq$ z $\leq$ 0.9, with stellar masses from 10$^{10}$ to 10$^{12}$ M$_{\odot}$. They found that the most massive galaxies at z $\sim$ 1 are already passive. In addition, they observed that lower mass galaxies have more extended SFHs, while the lowest-mass galaxies in their sample are actively forming stars. This is consistent with studies in the Local Universe, where the SFR of more massive galaxies tend to peak earlier in cosmic time \citep{2004Natur.428..625H, 2005ApJ...621..673T}. \cite{2016ApJ...832...79P} used 845 quiescent galaxies at 0.2 $\leq$ z $\leq$ 2.1 and computed the median SFHs of the sample in bins of stellar mass and redshift. They find that independently of the redshift and mass, the median SFHs rise, peak, and  decline until quiescence is reached. The duration of these phases depends on the redshift and stellar mass.

The Large Early Galaxy Astrophysics Census (LEGA-C) survey \citep{2016ApJS..223...29V, {2018ApJS..239...27S}, {2021ApJS..256...44V}} has provided a window at around z = 0.6 - 1 for studying galaxy properties. \cite{2019ApJ...877...48C} reconstructed the SFHs of a sample of quiescent galaxies at z = 0.6-1, using LEGA-C data. They identified secondary star-forming episodes (SFEs) that rejuvenate the average stellar populations of galaxies, moving them back to the star-forming main sequence (blue cloud) after an initial period of quiescence. Using LEGA-C and SAMI Galaxy data, \cite{2022MNRAS.512.3828B} discussed the evolution in the age-$M$/$R_{e}^{2}$ relation with redshift, studying populations of quiescent galaxies across 6 billion years (6 Gyr). They found that at high-z the gravitational potential correlates with the metallicity of galaxies; the deeper the potential wells, the more metal-rich the galaxies. However, they found no relation between age and surface density ($M$/$R_{e}^{2}$) at 0.60 $\leq$ z $\leq$ 0.76, in contrast with the observations at z $\sim$ 0. This evolution in the age-$M$/$R_{e}^{2}$ relation with redshift implies a change of star-forming and quiescent galaxies mass-size relations. This is consistent with galaxies forming more compactly at higher redshifts \citep{2014ApJ...788...28V, 2024MNRAS.527.6110O} and remaining compact during their evolution \citep{2022MNRAS.512.3828B}. From a sample of 3200 galaxies with 0.6 $\leq$ z $\leq$ 1, \cite{2023MNRAS.526.3273C} confirmed the main trends of global ages and metallicities with the stellar velocity dispersion in galaxies.

If we compare intermediate- and high-redshift (z $\sim$ 0.6-3) samples to nearby galaxies, we observe a range of differences. At fixed stellar mass, higher-z galaxies are more compact systems \citep{2004ApJ...600L.107F, 2014ApJ...788...28V, 2019ApJ...872L..13M}, more highly star-forming, have dynamically hotter turbulent star-forming gas \citep{2009ApJ...706.1364F, 2006ApJ...645.1062F, 2011MNRAS.417.2601W, 2012A&A...539A..92E} and higher molecular mass fraction \citep{2013ApJ...768...74T, 2015MNRAS.454.3792M}. At fixed stellar mass, star-forming galaxies are larger than their quiescent counterparts \citep{2011ApJ...742...96W, 2014ApJ...788...28V, 2017ApJ...838...19W}. The differences between the z = 0 star-forming and quiescent populations \citep{2021MNRAS.505.2247C} suggest a redshift dependence on galaxy evolution. At z = 2.1, \citet{2015ApJ...806....3A} analyzed the differences between field and cluster galaxies. They found that star-forming galaxies in clusters appear $\sim$20$\%$ redder than star-forming field galaxies at all masses. However, they did not find differences between field and cluster quiescent galaxies, indicating that the environment has not yet strongly influenced their evolution at z $\sim$ 2. \citet{2016ApJ...826...60A} also found that the environment does not have an impact on quiescent galaxies at z $\sim$ 1; however, in star-forming galaxies, their properties are indeed affected. 

At even higher redshifts (z $\sim$ 2-3), studies of individual objects or small samples have been mostly focussed on extracting the stellar mass and studying the initial mass function behaviour \citep{2017MNRAS.468.3071N, 2021ApJ...908L..35E, 2022ApJ...938..109F}. Most studies of distant star-forming galaxies have still had to rely on photometry alone for studying the stellar component. Instruments such as Keck-MOSFIRE, VLT-XSHOOTER and, JWST-NIRSpec/NIRCam have enabled studies of galaxies at these redshifts \citep{2014ApJ...788L..29B, 2018A&A...618A..85S, 2019ApJ...880L..31K, 2024ApJ...968....4P, 2024ApJ...963..129M} that resulted in the discovery of little red dots (LRDs). These are galaxies with significant stellar masses (median value $\sim$ Log$_{10}$(M$_{\star}$/M$_{\odot})$ = 9.4 at z $\sim$ 7) and old-and-massive quiescent stellar populations \citep{2017Natur.544...71G,2023Natur.619..716C,2024Natur.628..277G,2024arXiv241211861R,2025ApJ...979..249B, 2025NatAs...9..280D,2025ApJ...983...11W} for their redshifts.

In this paper, using different MUSE surveys probing different environments and redshift ranges (z = 0.1-0.9), we derive the stellar population parameters of a sample of $\sim$400 galaxies. To reconstruct the SFH of the sample, we analysed the variation in the  SFR as a function of cosmic time and identify their SFEs.

The paper is structured as follows. In Section \ref{sec:sample}, we describe the galaxy sample and how we built it. In Section \ref{sec:method}, we describe the full-spectrum method used to derive the stellar parameters and the SFH. Section \ref{sec:results} presents our results and discussion. Lastly, in Section \ref{sec:conclusions}, we provide a summary of our conclusions. Throughout the paper, we adopt a flat Universe, H$_{0}$ = 70 km s$^{-1}$ Mpc$^{-1}$, $\Omega_{\Lambda}$ = 0.7 cosmology.


    \begin{figure}
    \centering
       \includegraphics[width=9.8cm]{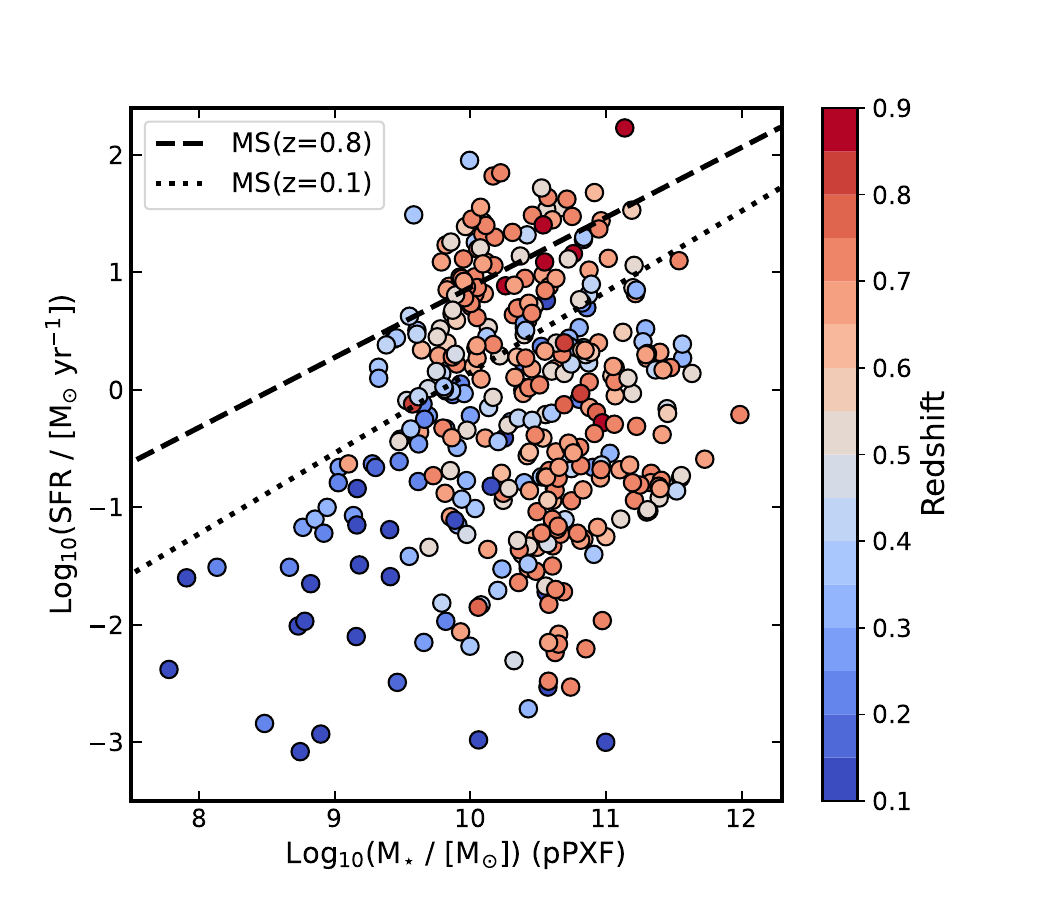}
         \caption{General properties of the galaxy sample. SFR as a function of the stellar mass (derived from the MUSE spectra) for the galaxy sample, colour-coded by the spectroscopic redshift. The dotted and dashed grey lines indicate the main sequence (MS) for an age of the Universe, respectively, of 12.4 Gyr (z = 0.1) and 6.8 Gyr (z = 0.8), given by Eq. (1) of \cite{2017ApJ...838...19W}.}
         \label{fig:sfr_mass_red}
    \end{figure}

\section{Galaxy sample}\label{sec:sample}

This work relies on data taken from four main different spectroscopic surveys. The spectroscopic surveys are located in the CANDELS/GOODS-S and CANDELS/COSMOS regions. MUSE-Wide \citep{2019A&A...624A.141U} covers $100 \times 1$ arcmin$^{2}$ fields each observed for one hour, while MUSE-Deep (HUDF, \citealp{2017A&A...608A...1B, 2023A&A...670A...4B}) covers 9 x 1 arcmin$^{2}$ mosaic for 10 hours, $1 \times 1$ arcmin$^{2}$ field to 31 hours \citep{2006AJ....132.1729B}, and 141 hours on a circular field with $1^{\prime}$ diameter \citep[MXDF;][]{2023A&A...670A...4B}. MAGIC survey \citep{2024A&A...683A.205E} targets 14 massive galaxy structures, mostly groups, having exposure times from 1 to 10 h. MUSE-Wide has a wide area of coverage, so that it includes massive galaxies. MUSE-Deep and MXDF go to lower masses and higher redshift, while MAGIC probes different environment, thereby increasing the number of massive galaxies expected for dense environments.

\subsection{Selection criteria}\label{sec:selection} 
    \begin{figure*}
    \centering
       \includegraphics[width=18cm]{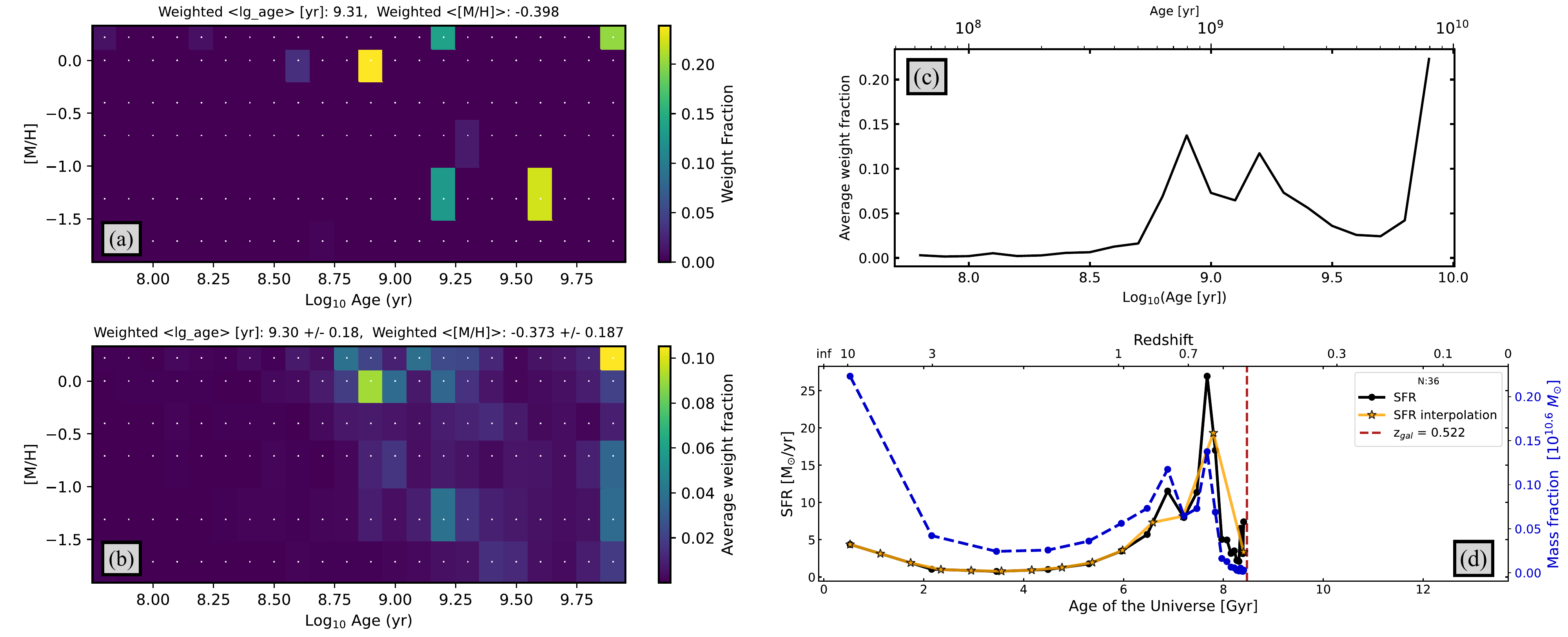}
         \caption{Example of the population analysis for one galaxy in the sample.
         \textit{Panels (a) and (b):} Distribution of the stellar pPXF weights, coloured by the weight fraction or bolometric luminosity of different stellar populations with given age and metallicity. Panel (a) corresponds to the unregularised fit. Panel (b) shows the averaged weights over 500 bootstraped spectra of the original spectrum. \textit{Panel (c):} Averaged weight fraction, based on weights of SSP models returned by the fitting, and related to the stellar mass, as a function of the stellar age. \textit{Panel (d):} SFR as a function of time, shown by the orange line. The SFR prior to smoothing and interpolation is shown by the black line. The dashed-blue line shows the stellar mass assembly with the mass fraction plotted on the right y-axis. The red dashed line shows the redshift of the system. The galaxy's catalog index ($N$) is shown in the legend of the plot.}
         \label{fig:example}
    \end{figure*}

The galaxy sample was selected based on each respective spectroscopic integrated signal-to-noise ratio (S/N). The main focus for building the sample was maximising the number of galaxies with high S/N to be analysed. Therefore, a primary run using the Penalized PiXel-Fitting method \cite[pPXF\footnote{Available in \href{https://pypi.org/project/ppxf/}{https://pypi.org/project/ppxf/}},][]{2004PASP..116..138C,2017MNRAS.466..798C}
was used to obtain a global estimator of the galaxy S$\slash$N (using the full wavelength range) for $\sim$3500 objects. 
Our criteria for building the sample are based on the integrated S/N of the galaxies. Hence, we extracted one spectrum per galaxy corresponding to the sum of the flux within 1 R$_{\rm eff}$. The effective radii were taken from the 3D-HST survey \citep{2014ApJS..214...24S} and COSMOS2015 \citep{2016ApJS..224...24L} photometric catalogs. We chose to use a set of 195 templates from the Indo-US stellar library \citep{2004ApJS..152..251V}. This library has a constant spectral resolution of 1.35 $\AA$ which is better than the average spectral resolution of MUSE data (2.5 $\AA$), corresponding to 2.27-1.31 $\AA$ at z = 0.1 and z = 0.9, respectively. We fit the stellar continuum and we only fit the absorption lines that did not overlap with any emission-line. We then masked all known emission-lines in the MUSE wavelength range, according to the galaxy's redshift. From this primary fit we derived the velocity dispersion ($\sigma$) of the sample. We adopted a threshold in the integrated S/N $\geq$ 5 within an aperture of 1 R$_{\rm eff}$ as a sample selection limit.

\subsection{Global properties}\label{sec:global-properties}

Based on the S/N cut-off, the sample ended up comprising 393 galaxies: 218 galaxies located within the MUSE-Wide footprint (i.e. MUSE-Wide, HUDF, MUSE-Deep and MXDF, see \citet{2019A&A...624A.141U}), with stellar masses ranging from $\sim$10$^{7.8}$ M$_{\odot}$ to 10$^{12} $ M$_{\odot}$; additionally, 175 galaxies are members of the MAGIC survey spanning from $\sim$10$^{9.3}$M$_{\odot}$ to 10$^{11.3}$M$_{\odot}$. The surveys cover in total an area of $\sim$126 arcmin$^{2}$ and target  low-to-intermediate-density environments (within the MUSE-Wide footprint), or  small to big groups (MAGIC footprint).\\

Our sample ranges from z$ \sim$0.1 to 0.9, covering around 6 Gyr in cosmic time. Around 50$\%$ of the galaxies have redshift 0.6 $\leq z\leq$ 0.9. Figure \ref{fig:sfr_mass_red} shows the stellar masses derived using pPXF when fitting MUSE and photometry data (see Section \ref{sec:emission_lines}) for the sample as a function of the  SFR derived from SED fitting methods. Henceforth, we  refer to the stellar masses derived using pPXF as spectral masses. The SFR for galaxies within the MUSE-Wide area was determined using the FAST code \citep{2009ApJ...700..221K} as part of the 3D-HST program \citep{2014ApJS..214...24S}. In the case of MAGIC galaxies, the SFR was computed using CIGALE code \citep{2019A&A...622A.103B} within an aperture of 3'' for each galaxy, as presented in \cite{2024A&A...683A.205E}. We see that our galaxy sample covers the star-forming main sequence (MS), given by Eq. (1) of \cite{2017ApJ...838...19W} for all the redshift ranges. We have actively star-forming, but also quenched, galaxies in all redshift ranges. However, the low-stellar-mass ($\leq$10$^{9}$ M$_{\odot}$) systems are predominately galaxies with low levels of star formation (see Section \ref{sec:bias} for more details on observational bias). In terms of SFR, both set of measurements (using CIGALE and FAST)  cover the quiescent and star-forming regions. However, from the stellar masses derived using SED fitting, we observe that galaxies within the MAGIC footprint sample are, on average, more massive (from 10$^{9}$ to 10$^{12}$ M$_{\odot}$) than in MUSE-Wide (10$^{7}$ - 10$^{12}$ M$_{\odot}$).

\section{Method}\label{sec:method}

We performed three different pPXF runs. The first one was done to build the galaxy sample, where we used Indo-US templates, masked emission lines, and retrieved the galaxy's S/N and $\sigma$ (see Section \ref{sec:selection}). In the second, we ran pPXF with E-MILES templates and masked emission lines, computing the stellar population parameters and SFH (Section \ref{sssec:sp} and \ref{sec:sfh_method}). Lastly, we used pPXF with E-MILES models, without masking the emission lines, but including   photometry bands to compute the stellar masses and characterise the emission lines (Section \ref{sec:emission_lines}). We present an example based on one galaxy, with all the fits performed, in Figure \ref{fig:example_36_appendix}.

\subsection{Stellar population parameters}\label{sssec:sp}

We extracted the stellar population parameters from the absorption-line spectra of the sample galaxies using pPXF, which reproduces the observed spectrum with a combination of templates from a simple stellar population library. We were able to measure global quantities, as the spectrum of each galaxy corresponds to the sum of the flux of the spaxels within 1 R$_{\rm eff}$.

Since a set of stellar population models are needed for the spectral fitting, we selected the E-MILES library \citep{2006MNRAS.371..703S, 2011A&A...532A..95F}. This library has a moderately high spectral resolution of FWHM = 2.5$\AA$, ranging from 3540$\AA$ to 8950$\AA$ (similarly to the MUSE data).

Before the fitting process starts for each galaxy, we cut the number of templates to be used according to the age of the Universe at the redshift of the galaxy. In this way, we restricted the oldest stellar population used when reproducing the galaxy spectrum to be as old as the Universe was at the redshift of the fitted galaxy. As we wanted to extract the stellar population parameters, we masked several gas emission lines: [OII] $\lambda \lambda$3726,3728, H$\delta$ $\lambda$4101, H$\gamma$ $\lambda$4340, H$\beta$ $\lambda$4861,  [OIII] $\lambda \lambda$4958,5006,  [OI] $\lambda$6300,  [NII] $\lambda$6548, 6583, H$\alpha$ $\lambda$6563, and [SII] $\lambda \lambda$6716, 6730. Given that some data was observed using adaptive optics (AO), we also masked the wavelength range (5800-5980$\AA$) corresponding to the laser dichroic when pertinent. We used multiplicative polynomials of 4th order and zero additive polynomials to adapt the continuum shape of the templates to the observed spectrum. 

    \begin{figure*}
    \sidecaption
       \includegraphics[width=12cm]{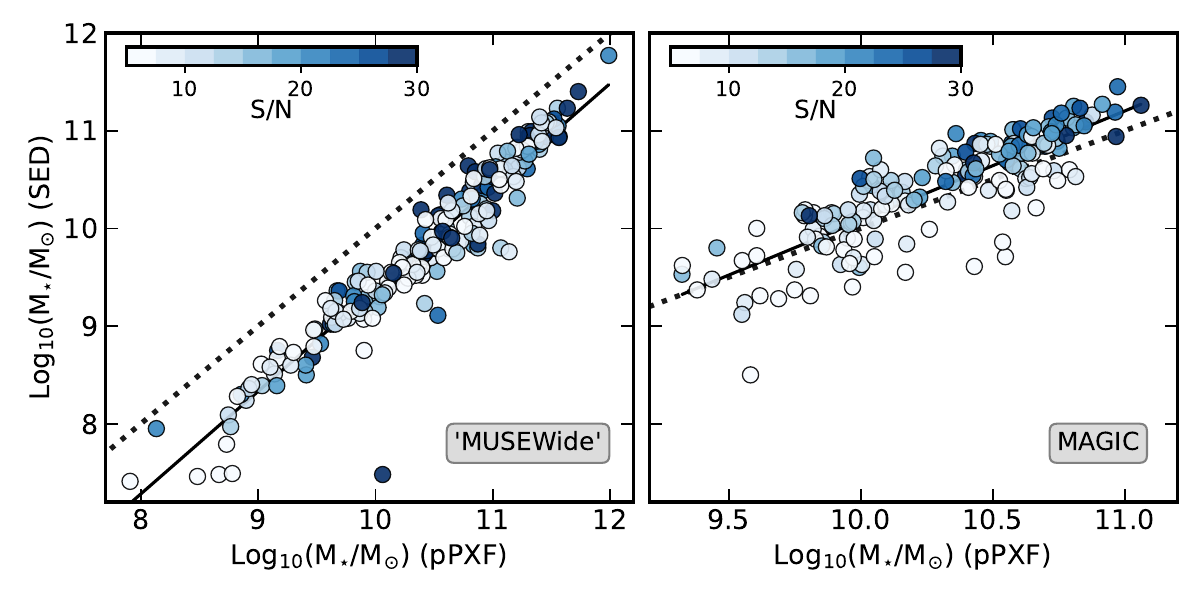}
         \caption{Comparison between the stellar masses derived using different methods and the pPXF stellar population mass for galaxies in the sample. Left panel:\ Galaxies within the MUSE-Wide footprint. Right panel: MAGIC galaxies. The dashed line is the one-to-one relation. The black-solid line is a linear fit to the derived masses. Symbols are coloured according to the integrated galaxy's S/N.}
         \label{fig:mass_comp}
    \end{figure*}
Due to the existing degeneracy between age, metallicity, and dust attenuation, we followed a similar method as \cite{2022A&A...659A.191E} when recovering the SFH of the sample galaxies. Instead of using the regularisation parameter in pPXF, we used a bootstrapping approach, which also allowed us to simultaneously get the error on the stellar population parameters (SPP). We performed an initial fit to the spectrum with the regularisation parameter set to zero. After getting the best-fitting model and the residuals for the galaxy spectrum, we perturbed the model spectrum with noise produced randomly within the range determined by the residuals, keeping  the zero regularisation at all times. We produced a total number of $N$ simulated spectra per galaxy, which we fitted again with pPXF without regularisation, obtaining their stellar population parameter (SPP). Thus, the average age and total metallicity of the galaxy are given by the average ages and metallicities over the $N$ = 500 iterations or simulated spectra, represented by the equations:

    \begin{equation}
     \langle \; Log_{10} \; (Age) \; \rangle \;=\; \frac{1}{N}\sum_{j=1}^{N} \left[ \frac{\sum_{i=1}^{n} Log_{10} \; (Age_{i}) \; w_{i}}{ \sum_{i=1}^{n} w_{i}} \right]_{j}
    \label{eq:age}
    ,\end{equation} 

    \begin{equation}
     \langle \; [M/H] \; \rangle \;=\; \frac{1}{N}\sum_{j=1}^{N} \left[ \frac{\sum_{i=1}^{n} [M/H]_{i} \; w_{i}}{ \sum_{i=1}^{n} w_{i}} \right]_{j}
    \label{eq:age}
    ,\end{equation} 

\noindent where $N$ is the number of simulated galaxy spectra, $n$ is the total number of templates used during the fitting of the $j$-$th$ spectrum. Here, $Age_{i}$ and $[M/H]_{i}$ correspond to the age and metallicity of the $i$-$th$ simple stellar population (SSP) model and $w_{i}$ is the weight assigned to the $i$-$th$ SSP model. We selected $N$ = 500 after running tests with different values of $N$ and getting convergent statistics. We derived the errors for $Age_{i}$ and $[M/H]_{i}$ from the standard deviation of the 500 measurements. In Figure \ref{fig:example}, panel (a) shows the original pPXF weight fraction distribution for one sample galaxy without regularisation; panel (b) corresponds to the weight distribution of the same system averaged over 500 simulated spectra.

\subsection{Gas fitting and stellar masses}\label{sec:emission_lines}

We also made use of the latest version of pPXF\footnote{pPXF version 9.1.1} \citep{2023MNRAS.526.3273C}, where photometry measurements in different bands can be used simultaneously with spectra during the fitting process to derive the stellar mass of the galaxies. We applied this updated method and derived the stellar masses of the galaxy sample. We used pPXF with E-MILES stellar templates again for the spectral fitting and derivation of the stellar masses. For galaxies in the MUSE-Wide footprint, we simultaneously fit a MUSE spectrum and more than than 30 bands ranging in wavelength from $\sim$0.2 $\mu$m to 4 $\mu$m. We used the photometric catalogs provided by the 3D-HST program, which include photometry of different instruments and wavelengths \citep[for more details:][]{2014ApJS..214...24S}. In the case of MAGIC galaxies, more than 20 photometric bands including HST/ACS were added to the MUSE spectra during the fit. We used a catalog provided by the MAGIC team \citep{2024A&A...683A.205E}. The catalogue contains measurements in different bands from the ultraviolet to the infrared.

From this run of pPXF, with E-MILES templates and all the photometric bands available, we fit the emission lines and derived the stellar masses. The emission lines were modelled in pPXF as Gaussians. We fit every Balmer line independently and kept them separate; thus, the individual lines have free fluxes and their flux ratios are not limited by atomic physics. Balmer lines and doublets have the same kinematics (velocity and velocity dispersion). These parameters are controlled by the keywords $tie\_balmer$ and $limit\_doublets$ which were set to 'false' in both cases. When calculating the total flux from doublets, we added up the individual fluxes and imposed the constraint that the ratio of strength between the two lines was between the corresponding atomic physics limits (0.35 - 1.5 in the case of [OII] doublet).

    \begin{figure*}
    \centering
       \includegraphics[width=17cm]{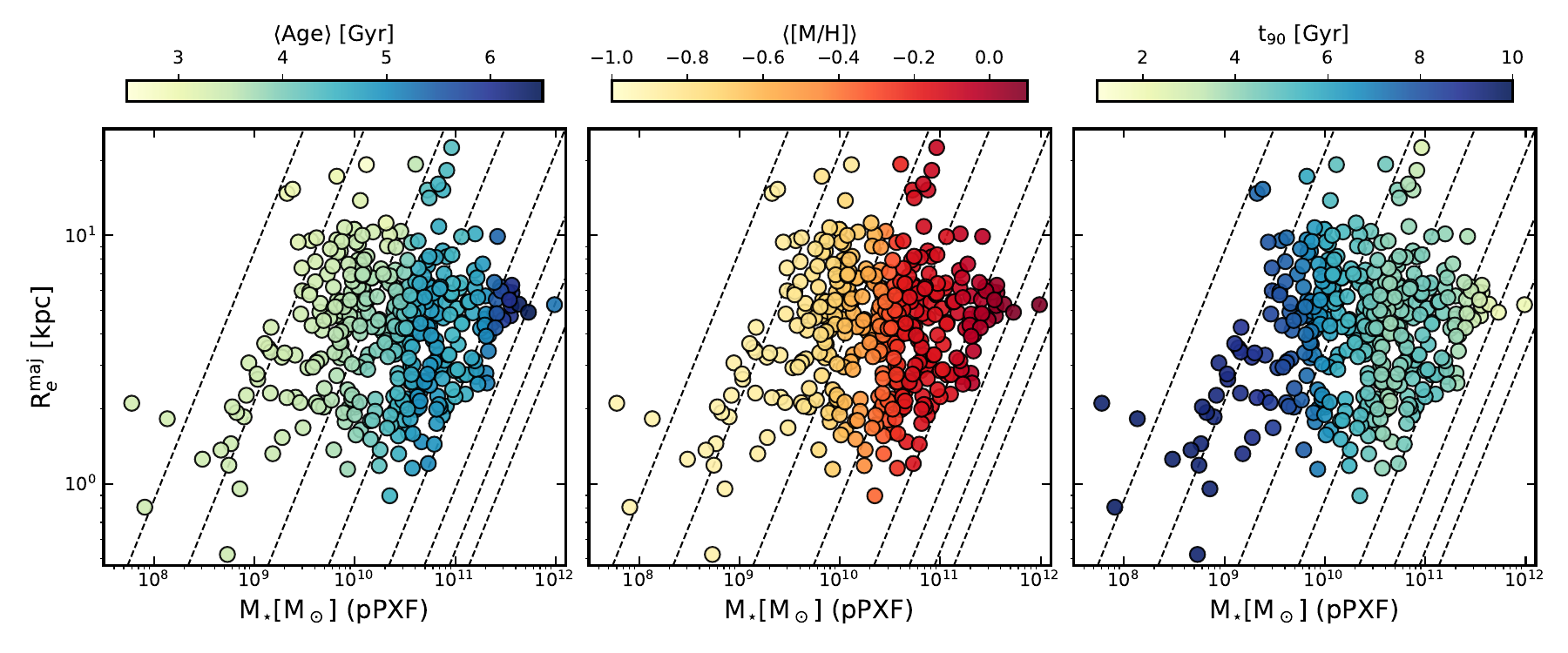}
          \caption{Plane of effective radius (R$^{\rm maj}_{\rm e}$) versus stellar mass coloured by SSP-equivalent population parameters measured within one effective radius. Colours of the left, central, and right panels correspond to the mean stellar age, mean stellar metallicity, and the time taken to form 90$\%$ of the star in the galaxy, respectively. These parameters were spatially averaged using the LOESS technique. Dashed-black lines show constant velocity dispersion of 10, 20, 50, 100, 200, 300, 400, 500 km s$^{-1}$ from left to right, derived by the virial mass estimator, M = 5R$^{\rm maj}_{\rm e}$ $\sigma^{2}$/G.}
         \label{fig:mass_size}
    \end{figure*}
In Figure \ref{fig:example_36_appendix}, panel (c)  presents an example of the photometry and spectrum fitting. In Figure \ref{fig:mass_comp} we present the derived stellar mass from the fitting process in comparison with stellar masses derived using SED fitting techniques, from \citet{2014ApJS..214...24S} in the case of galaxies within the MUSE-Wide footprint, using the FAST code \citep{2009ApJ...700..221K}  with a Chabrier initial mass function \cite[IMF,][]{2003PASP..115..763C}. In the case of MAGIC galaxies, the stellar masses were derived within an aperture of 3'' \citep{2024A&A...683A.205E}, using CIGALE with a Salpeter IMF \citep{1955ApJ...121..161S}. In the case of our stellar masses, we used E-MILES models with Universal Kroupa IMF \citep{2001MNRAS.322..231K}. We see that linear fits to the derived mass, for both the MUSE-Wide and MAGIC samples, have slopes (m) close to one (m = 1.05 and m = 1.12, respectively). The literature SED masses derived for the MAGIC sample agree better with ours, compared to those derived for MUSE-Wide, which are systematically higher than the ones derived using SED techniques. The differences in the computed stellar masses could be related to the different IMFs, stellar remnants, and gas loss used to derive them.

\subsection{Star formation history}\label{sec:sfh_method}

The SFH was computed from the resulting averaged weight fraction grids obtained during the spectral fitting with E-MILES templates and masked-emission lines. The normalised weight distribution corresponds to the V-band light-weighted fractional contribution of a stellar population with certain age and metallicity to the overall galaxy spectrum. Using the age-metallicity grid (see Figure \ref{fig:example}, panel b), the SFH is derived by collapsing the metallicity axis (y-axis), resulting in a sum of weights with different ages (see Figure \ref{fig:example}, panel c); in other words, we take the sum of the stellar populations with different metallicities, but similar ages. 

The SFH of a galaxy represents how its stellar mass was assembled over cosmic time. From the fitting, we obtained the age distribution and the stellar mass fraction relative to the total stellar mass of the stellar populations of a galaxy. Thus, to go from stellar ages to cosmic time, we considered the redshift of the galaxy ($z$) and calculated the age of the Universe at $z$. Therefore, the time when a stellar population of age $Age_{i}$ was formed is given by
    \begin{equation}
     t_{i} = t_{z} - Age_{i}
    \label{eq:time}
    ,\end{equation} 
where $t_{z}$ is the age of the Universe at redshift, $z$, and $Age_{i}$ is the age of the $i$-th stellar population of a galaxy. In this way, we are able to reconstruct the history of the mass assembly of the galaxy. Thus, we can trace when in time the different stellar populations started forming and how fast the process occurred. Furthermore, the SFH could also contain stars that were not necessarily formed in situ and were accreted. However, there is no information about these stars in the stellar population model; thus, the in situ and ex situ star formation cannot be disentangled. In panel d of Figure \ref{fig:example}, we give an example of the SFH for one galaxy in the sample,  shown as the \text{blue-dashed} line.\\

\subsection{Star formation rates}\label{sec:sfr}

We estimated the SFRs of the galaxy sample using two different approaches: from the spectral emission lines, and from the SFHs of the galaxies. Firstly, the ideal emission line to compute instantaneous SFR of a galaxy is H$\alpha$ at $\lambda$6563. However, since our MUSE sample ranges from $z \sim$ 0.1-0.9, we are only limited to observing this line up to $z \lesssim$ 0.4 due to the instrument's wavelength coverage. With this assumption, we also use [OII] $\lambda$3727, 3729 doublet to trace the ionised gas, and from this line estimate the SFR. We follow \citet{2022A&A...665A..54M} to derive the SFR from the [OII] lines. The [OII] lines fall into the MUSE wavelength range from $z \gtrsim$ 0.3. Computing the SFR from the H$\alpha$ line and [OII] doublet allows us to probe  recent star formation in a galaxy, namely, around the last 10 Myr \citep{1998ARA&A..36..189K}.

Secondly, as we aim to study the extended SFR of the sample and not all galaxies have emission lines, we compute it using the integrated stellar spectral fitting. From the galaxy fit, we obtained the age of the stellar populations and their mass fractions. Using both distributions we can estimate how galaxies were assembling their stellar mass over time. Therefore, the SFR as a function of time is given by
    \begin{equation}
     SFR(t_{i}) = \frac{M_{\star}(t_{i})}{\Delta t_{i}}
    \label{eq:sfr}
    ,\end{equation} 
where M$_{\star}$(t$_{i}$) is the stellar mass assembled at time $t_{i}$ (see Equation \ref{eq:time}) and $\Delta t_{i}$ is the time resolution in linear space of the $i$-th template, which is given by the stellar population library.\\

As the stellar templates are not regularly sampling the time dimension (the coverage for younger stellar populations is more finely sampled than for older ones), an artificial rise in the SFR driven by small values of $\Delta t_{i}$ can occur. To remove such artifacts, we calculated the median SFR for the last 0.5 Gyr of every galaxy. Then, we interpolated the values using bins of 0.5 Gyr and computed the final SFR. To this end, we  used regular time-spacing and smoothed possible non-real peaks in the SFR due to the coverage of the templates. An example of the SFR as a function of cosmic time, prior to and after interpolation and smoothing is presented in panel (d) of Figure \ref{fig:example} with black and orange lines, respectively.
    \begin{figure*}
    \sidecaption
       \includegraphics[width=12.7cm]{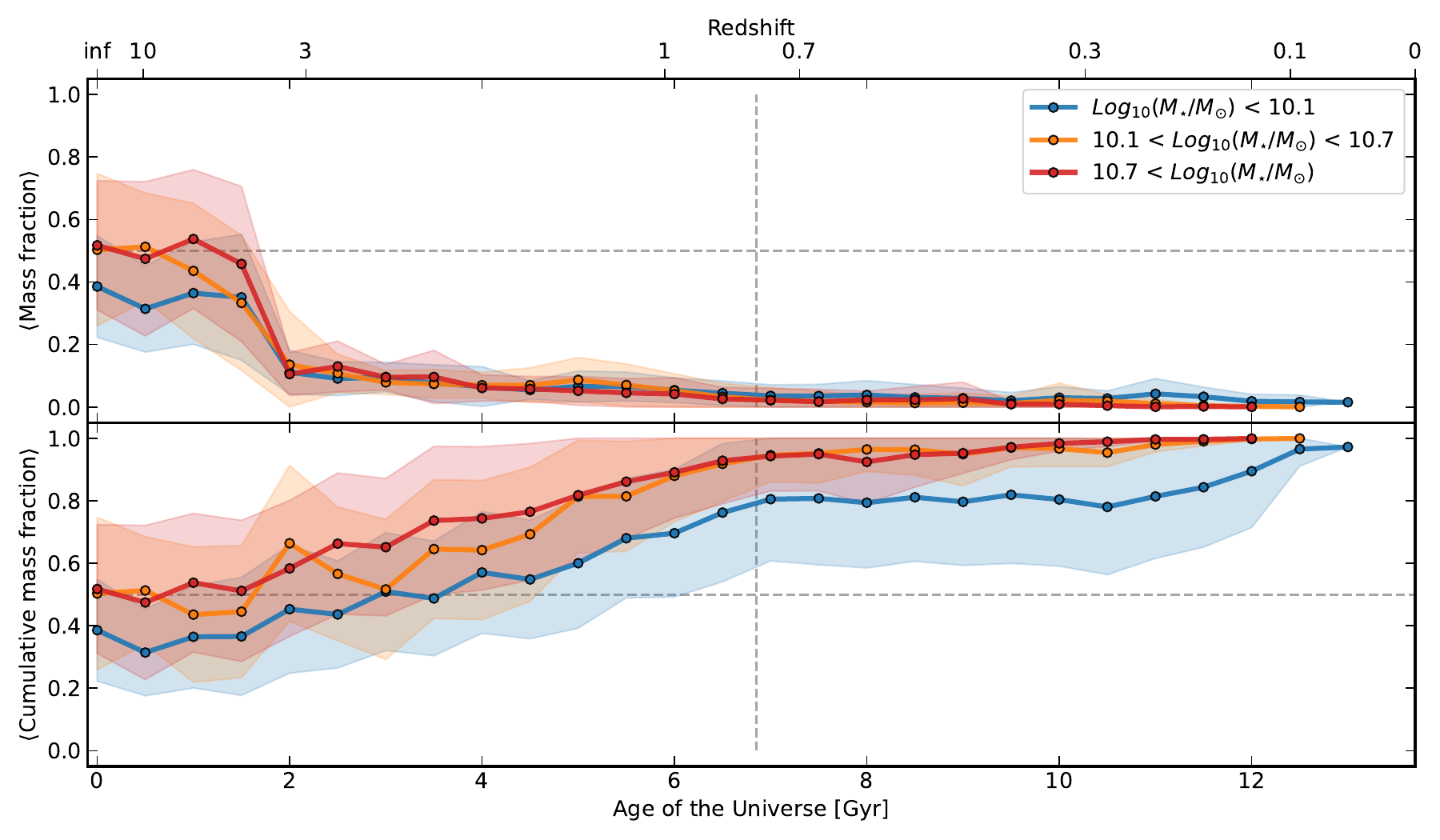}
         \caption{\textit{Top:} Mean mass fraction for the galaxy sample computed in bins of 0.5 Gyr. \textit{Bottom:} Mean cumulative mass fraction of the galaxy sample. In both panels, the colours indicate different stellar mass bins. Thick lines indicate the mean values and the surrounding shaded area the 1$\sigma$ dispersion of the distributions. We have 135 low-mass galaxies, 140 intermediate-mass, and 118 high-mass systems. Horizontal and vertical grey-dashed lines indicate the half-mass-and-cumulative fraction  and half-age of the Universe, respectively.}
         \label{fig:mf_cmf}
    \end{figure*}

\section{Results and discussion}\label{sec:results}

\subsection{Stellar population parameters}\label{sec:sp}

We present the stellar mass and physical galaxy size in Figure \ref{fig:mass_size}, as a function of different stellar population parameters. The size parameter $\rm R_{\rm e}^{maj}$, which corresponds to the major axis of the half-light isophote \citep{2010MNRAS.401.1099H}, was corrected by the galaxy's redshift scale to go from the angular to the physical size of the galaxies. In the case of galaxies within the MUSE-Wide footprint, the half-light isophotes correspond to a combination of the F125W + F140W + F160W band images taken from the catalog of the 3D-HST survey. In the case of the MAGIC galaxies, the half-light isophotes result of a sum of YJHK$_{s}$ and z$^{++}$ images from the COSMOS2015 catalog \citep{2016ApJS..224...24L}. The single stellar population (SSP) parameters were computed within 1 $\rm R_{\rm eff}$ and are shown in the panels with different colour scales smoothed using LOESS technique \citep{Cleveland1988LocallyWR} implemented by \cite{2013MNRAS.432.1862C}. With dashed lines, we show lines of constant velocity dispersion, derived from the virial mass estimator M = 5R$^{\rm maj}_{\rm e}$ $\sigma^{2}$/G \citep{2006MNRAS.366.1126C}.

We find that the mean stellar ages of galaxies follow a trend with constant velocity dispersion (left panel of Figure \ref{fig:mass_size}). The older the galaxies are, the higher their velocity dispersion is. At fixed stellar mass, more compact systems are on average older. We also observe in the mass-size plane, a correlation between the stellar mass and mean stellar metallicity. More massive galaxies are more metal-rich almost independently of their physical size. There is also a tendency for high-velocity dispersion systems to have more metals. Similar trends have been observed, independently of the environmental density, in the Local Universe \citep{2005ApJ...621..673T,2010MNRAS.404.1775T, 2015MNRAS.448.3484M}. In the right panel of Figure \ref{fig:mass_size} we present the mass-size plane coloured by the formation timescales, shown by t$_{90}$ or the time needed to assemble 90$\%$ of the galaxy's stellar mass. It shows that more massive galaxies form their stellar mass faster than lower-mass systems. We also observe that high-velocity-dispersion systems tend to have shorter formation timescale. These results agree with previous ones from nearby galaxy samples using SDSS stacked spectra \citep{2009ApJ...698.1232V, 2010MNRAS.405..948S}. Hence, the stellar age changes with the velocity dispersion, while the metallicity and t$_{90}$ change with stellar mass. In Appendix \ref{fig:mass_size_raw}, we present the non-smoothed version of Figure \ref{fig:mass_size}.

Several surveys in the Local Universe have studied the mass-size plane. \cite{2015MNRAS.448.3484M}, using early-type galaxies (ETGs) from the ATLAS$^{3D}$ observed that at fixed mass, compact systems are older, have more metal, and are more $\alpha$-enhanced than larger counterparts. \citet{2023A&A...673A.147P} observed that more massive late-type galaxies (LTGs) are older and more metal-rich. Notwithstanding, the galaxy's morphology was not a criterion to build our sample, we find similar results regarding the mass and stellar ages.

\cite{2018MNRAS.476.1765L} studied $\sim$ 2000 early-type and spiral galaxies from the MaNGA survey and concluded that these systems occupy different regions in (M$_{\star}$, R$_{e}^{maj}$) plane. The stellar population properties of these galaxies vary systematically, where high-velocity dispersion systems are older, exhibit larger stellar mass-to-light ratios, and contain more metal. Therefore, since $\sigma$ is a proxy for the bulge fraction when the stellar population of galaxies evolves their bulge fraction also increases. We observe a similar behavior, where older galaxies are more velocity dispersion-dominated systems. Nevertheless, the $\sigma$ and metallicity correlation is weaker in our sample.

With a sample of $\sim$1300 galaxies of all morphologies and environments from the SAMI survey, \citet{2017MNRAS.472.2833S} observed that galaxies with high stellar surface mass density have older stellar populations, more metals, and are more $\alpha$-enhanced than less dense galaxies.

There is a link between the SFH and abundance ratio of elements such as iron and $\alpha$-process elements \citep{2011MNRAS.418L..74D}. If we assume instantaneous recycling, there is a connection between [$\alpha$/Fe] and the star formation timescale, and therefore the observed metal build-up of galaxies. This relies on the assumption that stars are chemically enriched by progenitor generations (e.g. supernovae Type Ia, core-collapse supernovae), which enrich the stellar population of galaxies with different elements over various timescales. 
High $\alpha$-abundance points towards a short star formation timescale (ignoring the IMF). On the other hand, [$\alpha$/Fe] traces SF of $\leq$ 1 Gyr. 

At z $\sim$ 0.8, a similar trend as in the nearby Universe is observed. Using LEGA-C galaxies, \cite{2023MNRAS.526.3273C} found that both metallicities and ages follow similar lines of constant stellar velocity dispersion or equivalently lines where M$_{\star} \propto$ R$_{e}^{maj}$. Similarly to lower redshift, quiescent galaxies with stronger gravitational potential are more metal-rich \citep{2022MNRAS.512.3828B, 2025A&A...695A..86N}. In this context, our sample shows that some galaxy correlations observed in the low-z studies as age-$\sigma$, mass-metallicity, and mass-t$_{90}$ are already placed at $\sim$z = 0.8 without targeting specific galaxy types or environment (see Section \ref{sec:bias} for more details). For a complete version of stellar population parameters for the sample, we refer to Table \ref{tab:sample_table}. 
    \begin{figure*}
    \sidecaption
       \includegraphics[width=12.7cm]{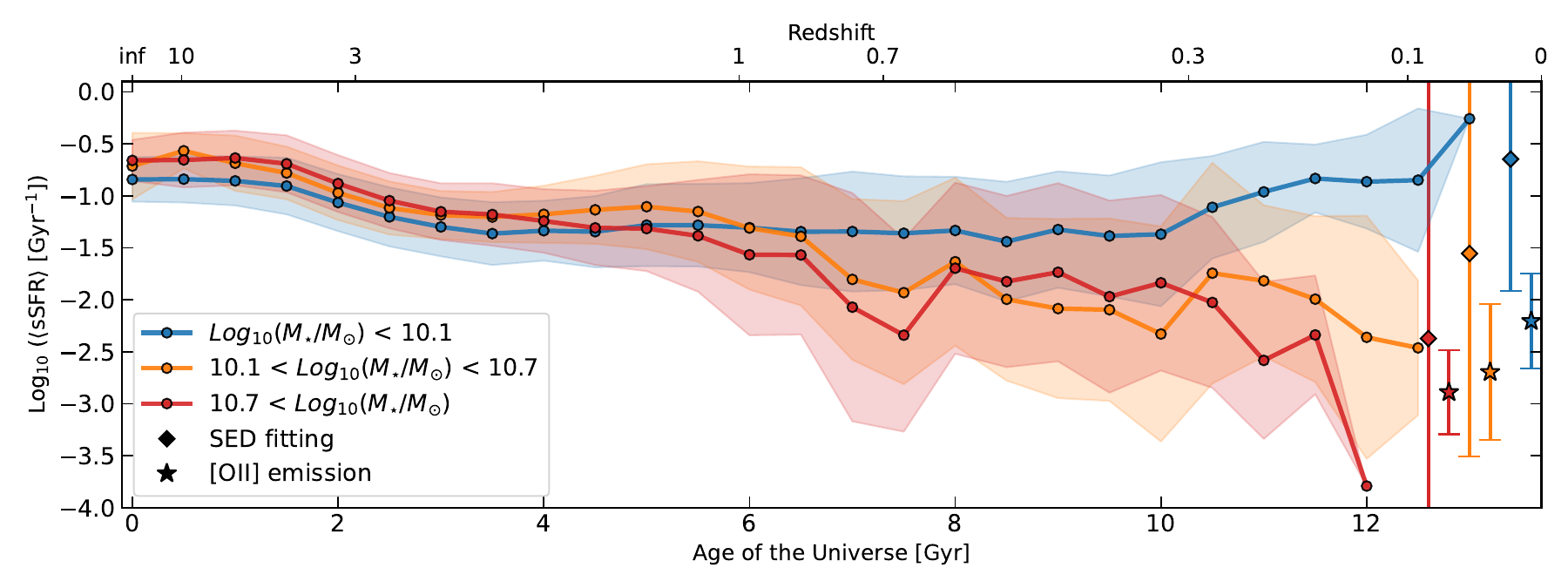}
          \caption{sSFR for the galaxy sample, averaged over three mass bins, indicated by the legend, as a function of cosmic time. Thick lines correspond to the mean values and the surrounding shaded area to the 1$\sigma$ dispersion of the distributions. Diamonds and stars indicate the mean sSFR from SED and [OII] emission line measurements, respectively, placed in arbitrary positions in the x-axis for better visualisation.}
         \label{fig:ssfr}
    \end{figure*}
\subsection{Stellar mass assembly}\label{sec:sfh}

In this section, we present the SFH of our galaxy sample. In Figure \ref{fig:mf_cmf} we show the resulting SFH plots for the galaxy sample separated in three different stellar mass bins, from low (Log$_{10}$ (M$_{\star}$/M$_{\odot}$) < 10.1) to high (10.7 < Log$_{10}$ (M$_{\star}$/M$_{\odot}$)) stellar masses. Each mass bin contains the distribution of SFHs, showing the mean behavior of the distribution with thick lines and 1$\sigma$ dispersion with shaded area. The mass limits were selected to have a similar number of galaxies in each bin. Hence, we have 135 low-mass galaxies, 140 intermediate-mass, and 118 high-mass systems. We observed that both, low and high-mass galaxies assembled a high fraction of their stellar mass ($\sim$50$\%$) at early stages of the Universe, which are beyond our ability to trace accurately. At these very early stages, the rate at which they form stars depends on their final stellar mass: the more massive the galaxy is, the faster the mass assembly is. From z$\sim$10 to z$\sim$3, they keep assembling their stellar mass at the same rate. After z$\sim$3, there is a drop in the mass fraction that is assembled to less than $\sim$10$\%$ of the total mass. This implies that the star formation of galaxies is more efficient when the Universe is young, which may be related to the availability of pristine gas for star formation \citep{2013ApJ...765..104B, 2016ApJ...832....7A}. At redshifts z $\geq$ 1, it is also expected that most galaxies are gas-rich and have high SFRs as this redshift coincides with the peak of SFR density at around $z \approx$ 1.9 or $\sim$3.5 Gyr after the Big Bang \citep{2014ARA&A..52..415M, 2020MNRAS.498.5581B}. This behaviour of the cosmic SFH flows naturally from our work without any assumption of the SFH parametrisation in pPXF. From the lower panel of Figure \ref{fig:mf_cmf}, we can also see that, on average, as high-mass galaxies assemble faster their stellar mass, they reach quiescence earlier or have lower levels of star formation than low-mass systems.

\subsection{Specific star formation rates}

We calculate the specific star formation rate (sSFR = SFR/M$_{\star}$) as a function of time from the distribution of weight fractions provided by the spectral fitting. As the weight distribution shows the relative contribution of the different stellar populations to the total light, we can trace when the stellar populations were formed and reconstruct the behavior of their SFR throughout cosmic time (for more details see Section \ref{sec:sfr}).

In Figure \ref{fig:ssfr} we present the mean sSFR as a function of cosmic time, for the three different stellar mass bins. The SFR comes from Equation \ref{eq:sfr}, where M$_{\star}$ is the total mass (at the epoch of observations). We can observe that during the first 6 Gyr or until around z = 1, the levels of star formation per unit of stellar mass are similar for all galaxies. This could be interpreted as that the relative efficiency in converting gas to star is similar and independent of the final stellar mass of the systems, at these redshifts. After the first 6 Gyr, as more massive systems have assembled more stellar mass, their levels of star formation decrease in comparison with low-mass galaxies. Galaxies in the lowest mass bin reach, on average, their peak in SFR towards z = 0, marking them as very different from the rest of the population, which tend to stop, or significantly reduce, forming stars at that epoch. This behaviour gives clues that the mass assembly in high-mass galaxies is faster and they reach quiescence earlier than younger systems. In addition, we include the mean sSFR computed from SED fitting \cite{2014ApJS..214...24S, 2012ApJS..200...13B, 2016ApJS..224...24L, 2022ApJS..258...11W}, plotted at the right side of Figure \ref{fig:ssfr} for better visualisation. These measurements correspond to the most recent star-forming activity of the galaxies, generally the last 100 Myr \citep{2009ApJ...700..221K}. We also include the sSFR derived from the fitting of the [OII] doublet emission lines, which trace the SFR in the last $\sim$10 Myr. We note that not all the galaxies in our sample have emission lines in their spectrum. Thus, these measurements correspond to a sub-sample of galaxies separated in different mass bins. Nevertheless, we see that these values are in line with the sSFR calculated from the spectral fitting, with an indication of a decreased most recent SFR for the low mass galaxies.

\begin{figure}
    \centering
       \includegraphics[width=9.4cm]{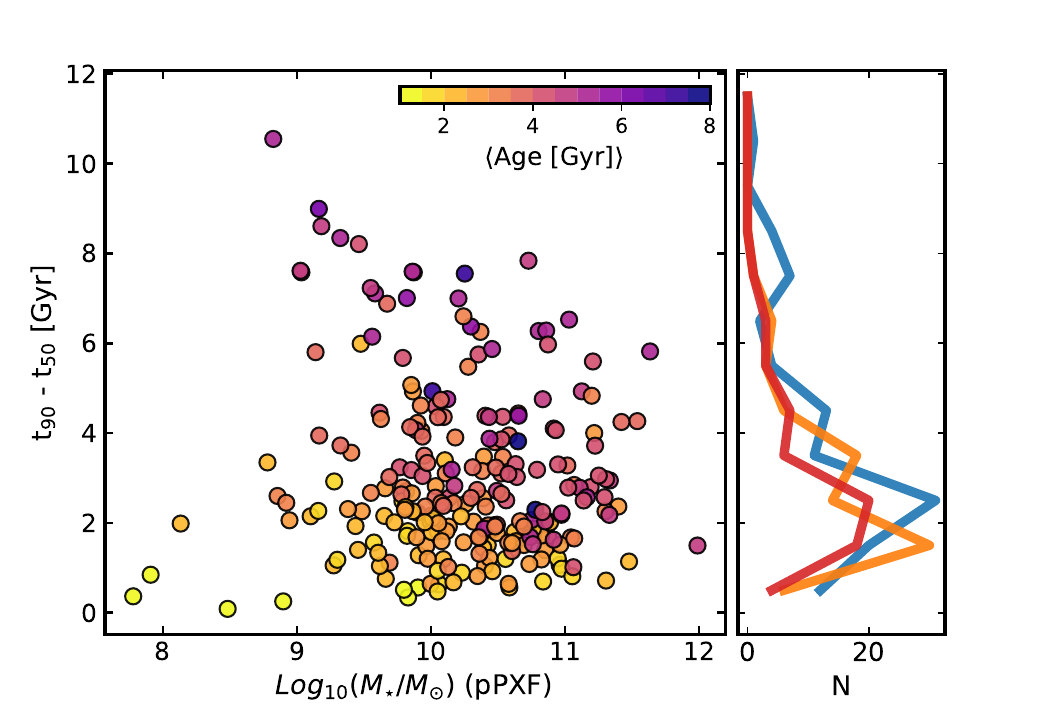}
         \caption{Time taken to form from 50$\%$ to 90$\%$ of the galaxy's stellar mass as a function of their final spectral stellar mass (derived with pPXF), coloured by the mean age of the galaxies. Right panel:\ Number of galaxies in bins of 1 Gyr for three different mass bins: low Log$_{10}$ (M$_{\star}$/M$_{\odot}$) < 10.1 in blue, intermediate 10.1 < Log$_{10}$ (M$_{\star}$/M$_{\odot}$) < 10.7 in orange, and high 10.7 > Log$_{10}$ (M$_{\star}$/M$_{\odot}$) in red.}
         \label{fig:t90}
    \end{figure}

Figure \ref{fig:t90} shows the time difference needed for the galaxies to assembly the 50$\%$ and 90$\%$ of their stellar mass. We observe the following: the majority of intermediate- and high-mass galaxies (> 10$^{10.1}$M$_{\odot}$) form the last 40$\%$ of their stellar mass in less than $\sim$4-5 Gyr. Some galaxies form those 40$\%$ in a longer period of time, hence, their stellar light is dominated by older stellar populations (darker colours). However, lower-mass systems show a somewhat larger scatter, as shown by the blue-line histogram in the figure. At stellar masses < 10$^{9.5}$M$_{\odot}$, galaxies are at z < 0.3 and have low levels of star formation in the last 10 Myr (see Figure \ref{fig:sfr_mass_red}). We observe that for these dwarf systems, the mass assembly can either be slow (> 6 Gyr), indicating that they have mostly old stellar populations, or they can be fast (in less than $\sim$2 Gyr) at low redshift, so their light is dominated by younger stellar populations. This scenario is in agreement with the change in the slope of the mean sSFR for low-mass galaxies at z > 0.3 (Figure \ref{fig:ssfr}). 

We also observe in Figure \ref{fig:ssfr} that high-mass galaxies decrease in their rate of star formation with time, while low-mass galaxies on average reach their peak SFR at later stages. This is a similar result with respect to the shape of the mean cumulative mass fraction shown in the lower panel of Figure \ref{fig:mf_cmf}.

We estimated the fraction of active galactic nuclei (AGNs) in our sample as well. Therefore, we cross-matched our sample with X-ray data from the CHANDRA deep field-south \citep{2017ApJS..228....2L} and Karl G. Jansky Very Large Array (VLA) at 3 GHz (10 cm) \citep{2017A&A...602A...1S} surveys, in the CANDELS-S and COSMOS fields, respectively. Of the 218 galaxies within the MUSE-Wide footprint, 14 systems are classified as AGNs. In the case of galaxies in the MAGIC footprint, 11 out of 175 correspond to AGNs. In total, these systems correspond to $\sim$6$\%$ of our galaxy sample, and they do not bias the results.

\subsection{Star-forming episodes}\label{sec:sfe}

One of the main drivers in the evolution of galaxies are the changes in their levels of star formation. Using the technique described in Section \ref{sec:sfr}, we derived the SFRs of the galaxy sample as a function of time. These describe the different levels of star formation galaxies experience during their life. After deriving the smoothed SFR for the sample (see Figure \ref{fig:example} \textit{panel (d)}), we assume that the SFH of a galaxy can be modelled by a exponentially decaying SFR function plus a set of Gaussian peaks. We fit an exponential and a set Gaussian functions to the smoothed SFR using a least-squared fitting method. With this method we identify and describe different SFEs from the overall shape of the galaxy SFR, a similar approach used by \cite{2024ApJ...968..115W}. The selection of these fitting functions is due to the shape of the different SFR for the sample galaxies and, also due to their symmetric and asymmetric properties. Therefore, the observed SFR for a galaxy is given by
    \begin{equation}
     SFR(t) = \sum ^{N}_{i=1} SFE_{i}(t)
    \label{eq:sfe}
    ,\end{equation} 
where $t$ is the cosmic time since the Big Bang and, SFE$_{i}$ are the $N$ different SFEs that could be described as Gaussian or exponential distributions.

\begin{figure}
    \centering
       \includegraphics[width=9cm]{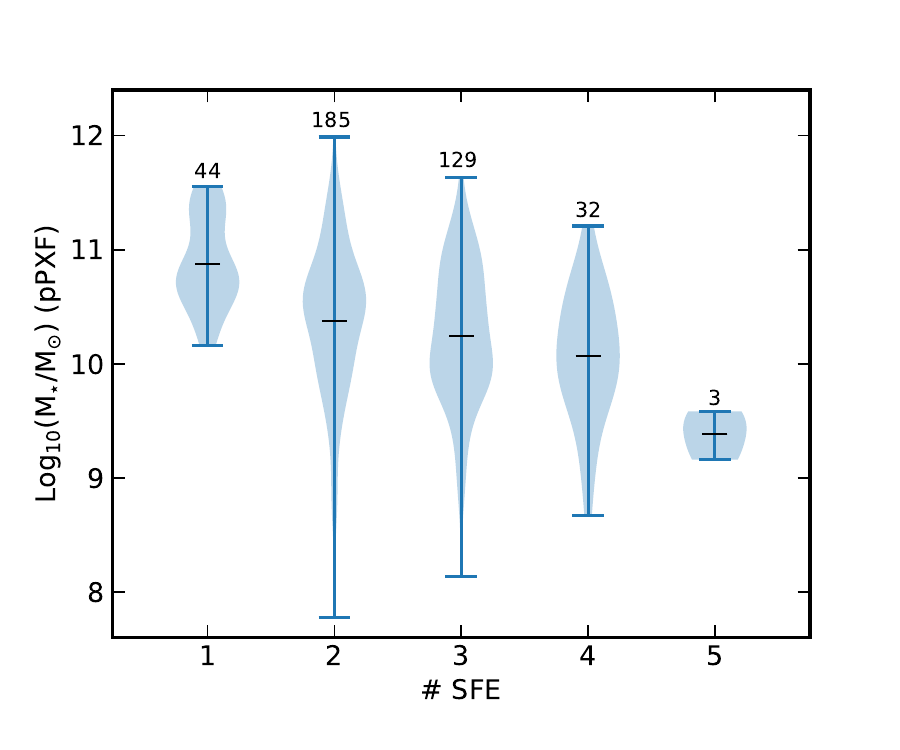}
         \caption{Number of SFEs as a function of the stellar mass. The violins show the mass distribution with the number of galaxies displayed at the top of each set. Black lines correspond to the mean of each distribution. The majority of galaxies with one SFE have SFR described by one exponential function. Higher numbers of SFEs are mostly modelled by exponential and Gaussian distributions. In Appendix \ref{fig:sfe_example} we present some examples of galaxies with different SFEs.}
         \label{fig:sfe}
    \end{figure}

We found that more than 85$\%$ of the galaxies in the sample have more than one main episode of star formation with the highest number of SFE in our data is five. We show the distribution of SFE in Figure \ref{fig:sfe}. All galaxies have an early SFE marking the initial star-forming event and making the oldest stars. This initial SFE is typically recovered by our method with an exponential decaying SFR, implying that we do not have an ability to resolve the onset of the primary SF and the SF within the first 1 Gyr in general. There is a weak correlation between the stellar mass and the total number of SFE. Higher-stellar-mass galaxies are more likely to have one or two main episodes of star formation, which are normally described by the combination of an exponential plus a Gaussian. The most massive galaxies in the sample typically have one main SFE, which is described by an exponential function. On the other hand, low-mass galaxies tend to be the result of more SFE, but the SF main event is usually described by an exponential distribution at the beginning of their history of mass assembly, accounting for up to 50$\%$ of their mass, followed by various Gaussian-like SFE at later stages.  
This suggests that more massive galaxies have one or two main events where they accumulate great fractions of their total stellar mass. Therefore, they consume their gas supplies rapidly and more efficiently, reaching quiescence faster. This scenario is in line with the results on the cumulative mass fractions shown in Figure \ref{fig:mf_cmf} and the sSFR results in Figure \ref{fig:ssfr}. 

\begin{figure}
    \centering
       \includegraphics[width=9.3cm]{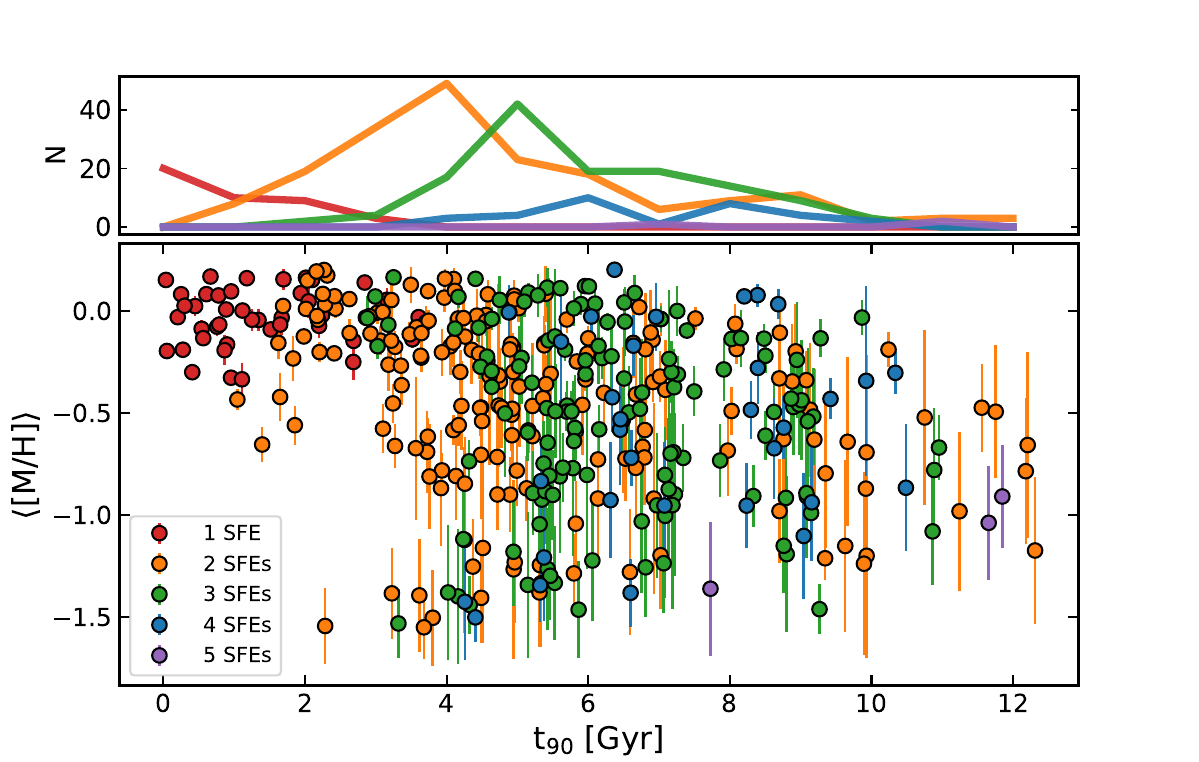}
         \caption{Mean galaxy metallicity as a function of t$_{90}$ coloured by the number of SFEs, ranging from 1 to 5. Top panel shows the distribution of SFEs as a function of the galaxy formation timescale, given by t$_{90}$.}
         \label{fig:sfe_metal}
    \end{figure}

In Figure \ref{fig:sfe_metal}, we present the mean metallicity as a function of t$_{90}$ coloured by the different number of SFEs. We see that galaxies that form 90$\%$ of their stellar mass very fast (< 3 Gyr) have high metallicities. These are the most massive galaxies in the sample, and their stars were formed mainly in an early intensive SFE (or a sequence of short bursts). However, metal-rich, but somewhat less massive, galaxies can also be the result of multiple SFEs, where the number of events is related to the formation timescale (t$_{90}$). This is related to the time resolution in which the stellar evolution processes work. In linear space, the SSP models provide a finer resolution for younger stellar populations. The variation with age ($\geq$ 5 Gyr old) in the continuum shape of the SSP spectra at fixed metallicity is almost indistinguishable. Therefore, if we have multiple SFEs occurring close together in time, especially when the Universe was young, we are not able to distinguish them individually. The spectra of the stellar population resulting from these SFEs at high-z are alike and detected as one main SFE. In Appendix \ref{fig:sfe_example} we present examples of galaxies with different numbers of SFEs. 

To investigate the level to which the spectral features are related to the number of SFEs, we stacked the spectra (Figure \ref{fig:stacked}) of galaxies with 1 and 4-5 SF events. In both samples, the number of stacked galaxies is similar ($\sim$35 systems). We observed various differences in the stacked spectrum of both groups. Firstly, galaxies with one main episode are, on average, at higher redshifts (z $\geq$ 0.5) than systems with 4/5 SFEs (z $\leq$ 0.5). Secondly, the stacked spectrum for galaxies with one event is dominated by 'old' stellar populations, mostly without the presence of emission lines. On the other hand, the light from galaxies with multiple episodes is dominated by 'young' stellar populations, showing the presence of gas emission lines (H$\beta$, [OIII], and [OII] mostly). We observed that multiple SFEs are found in galaxies with spectra that evoke rejuvenation events. \citet{2012A&A...539A..45L} noted that SDSS galaxies undergoing mergers have an excess of young stellar populations. In the context of gas-rich (wet) mergers, as there is abundant gas to support star formation and increase the SFR of the systems (e.g. \cite{2010ApJ...718.1158L,2011MNRAS.417..580P,2016ApJ...821...90A}), galaxies with a higher number of SFEs could trace merging activity. \citet{2019A&A...631A..87V} studied the evolution of major and minor mergers in a field that our sample covers as well, finding that the fraction of galaxies undergoing major merger at z $\sim$ 1 is around 10-20$\%$. In our sample, there are more than 85$\%$ of galaxies with SFE > 1, indicating that not all SFEs are necessarily associated with major mergers and that rejuvenating episodes might be triggered by either minor mergers or pure gas accretion.
\subsection{Stellar mass assembling and star-forming episodes}

In the context of the Local Universe, various studies \citep{1998MNRAS.295L..29K, 2000AJ....119.1645T, 2005ApJ...621..673T, 2010MNRAS.404.1775T, 2015MNRAS.448.3484M} have found that ETGs assembled high fractions of their stellar mass quickly and early in cosmic time. Similar results have been found for samples at z $\sim$ 1 \citep{2014ApJ...788...72G, 2015MNRAS.454.1332S}.
    \begin{figure*}
    \centering
       \includegraphics[width=15cm]{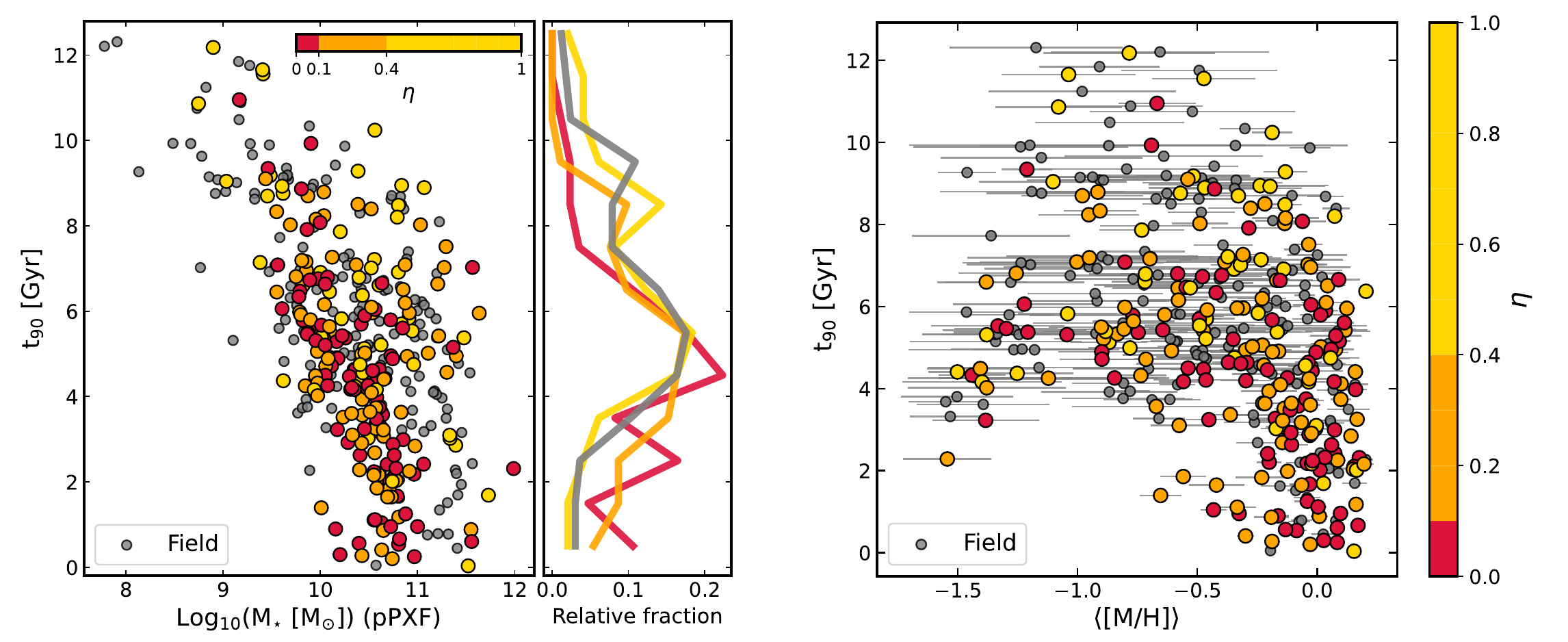}
         \caption{\textit{Left:} t$_{90}$ as a function of the stellar mass (derived from the spectra). The colours indicate different galaxy environments, quantified by the $\eta$ estimator. Grey dots correspond to galaxies that do not belong to groups or structures, thus denominated field galaxies. The right panel shows the relative fraction (respecting all galaxies in the same environmental category) of galaxies in different t$_{90}$ bins. \textit{Right:} t$_{90}$ as a function of the mean galaxy metallicity coloured by the environment.}
         \label{fig:environment}
    \end{figure*}

We observe in Figure \ref{fig:sfe} that most of the galaxies have more than one SFE. Moreover, the majority (> 95$\%$) of them have an early exponential SF event where they make their oldest stellar populations. The most massive galaxies in the sample are those that only had  an early exponentially decaying SFE, making their stellar mass early on and not having important SFE later on. In Figure \ref{fig:mf_cmf}, we observed that all galaxies made $\sim$50$\%$ of the stellar mass when the Universe was young ($\sim$2 Gyr). Thereby, the initial exponential SFE is the one making the oldest stars and contributing to a great fraction of their total galaxy stellar mass. Therefore, the initial SF event contributes to a high fraction of the total stellar mass of galaxies, and it occurred when the Universe was very young (z > 3). The remaining mass assembles slowly over a longer period of time.

Moreover, we can conclude that the most massive galaxies (> 10$^{10.7}$ M$_{\odot}$) in our sample, which are at z > 0.7, form early most of their mass typically in one exponential SF event. This result resonates with the findings made using the James Webb Space Telescope (JWST). \cite{2022ApJ...938L...2F, 2023ApJ...955...94F} found that JWST galaxies are much more evolved than expected at higher redshifts compared to HST observations. These high-z evolved galaxies have a wide diversity of morphologies (ETGs, disk galaxies, etc.) and stellar masses from $\sim$ 10$^{8.5}$ - 10$^{11}$ M$_{\odot}$ \citep{2023ApJ...946L..15K}. The presence of massive quenched galaxies at z > 3 \citep{2024arXiv241211861R} found by JWST agrees with our most massive galaxies forming a great fraction of their stellar mass when the Universe was less than 2 Gyr (or z > 3; see Figure \ref{fig:mf_cmf}). Similar results on the formation history have also been found by \citet{2024IAUS..377....3G}, \citet{2024MNRAS.534..325C} and \citet{2025ApJ...981...78N} with samples of massive and quiescent galaxies at 3 < z < 5. 

Our sample supports these JWST findings. We observe that the first exponential SFE for galaxies has a mean exponential decay of 2 $\pm$ 1.6 Gyr. Nevertheless, by z $\sim$ 2-3 a great part of the total stellar mass is assembled by the initial exponential SFE. The duration of the first SFE is, however, poorly constrained by our data and models. It is also possible that this first exponential SFE is actually made of multiple bursts of SF, much shorter even compared with the posterior SFE at subsequent epochs. The subsequent Gaussian events have different SF properties: they are less massive and have a shorter duration (mean Gaussian sigma of 0.77 $\pm$ 0.7 Gyr).

The duration of subsequent SFEs  sees significant results in quiescent galaxies at z $\sim$ 0.8 \citep{2022MNRAS.512.3828B}, where on average galaxies reach quiescence around z $\sim$ 1.2 and end up having a secondary SFE lasting $\sim$ 0.7 Gyr, contributing $\sim$2$\%$ to the total stellar mass of the galaxies. These systems could have analogues in our sample in the more massive galaxies with two SFEs (see Figure \ref{fig:sfe}) that after z = 1 reduce the stellar production but around z = 0.7 increase the SFR (see Figure \ref{fig:ssfr}). We also observe that the more massive galaxies in our sample at z = 0.8 have assembled on average more than $\sim$90$\%$ of their total stellar mass (Figure \ref{fig:mf_cmf}).

Looking at Figure \ref{fig:sfr_mass_red}, we see that the most massive galaxies in our sample cover both the star-forming main sequence and the quiescent part of the diagram. This SFR is tracing the galaxy's star formation state in the last $\sim$200 - 100 Myr. In Figure \ref{fig:ssfr}, we presented the SFR given by H$\alpha$ and [OII] emission features, which probe around the last $\sim$10 Myr. We observed that most massive galaxies in our sample also have low levels of recent star formation activity.

Our result regarding the fraction of the total stellar mass that is formed at z > 3 for galaxies in our sample resonates also with JWST results regarding LRDs. LRDs are examples of systems that are quenched by z$\sim$3 and therefore completed their (at least) first SFE. While these galaxies do not need to be considered very massive, they nevertheless have considerable mass, of a median $\sim$10$^{9.4}$ M$_{\odot}$ (Pérez-González et al. 2024; Matthee et al. 2024). Crucially, if we consider that those masses correspond to $\sim$50$\%$ of their total stellar mass, as in our sample galaxies form around half of their mass by z$\sim$3, their final stellar masses would be $\sim$10$^{9.7}$ M$_{\odot}$. Therefore, our sample should contain galaxies that were LRDs (at z > 3). As the subsequent evolution of LRDs could involve both a passive evolution or a set of rejuvenation episodes, it is possible that most of our galaxies actually passed through a LRD stage, where only a minority of such galaxies would remain quiescent down to the redshift range of our sample.

\subsection{The role of environment}\label{sec:env}

We would like to explore whether our results are influenced by the different environments our sample covers, specifically if there is a relation between the quenching of galaxies, the number of SFE and the environment. MUSE-Wide is a blind spectroscopic survey; therefore, the environment was not taken into account for the survey building. However, the MAGIC survey targeted groups of galaxies specifically at z $\sim$ 0.7, and one small cluster and it also contains galaxies in less dense environments at various redshifts. Therefore, we used a global density estimator to characterise the environment in our galaxy sample. We follow a similar approach as \citep{2024A&A...683A.205E}, where a friends-of-friends (FoF) algorithm was used to identify structures and galaxy groups. In addition, we consider the level to which galaxies are bound to their respective groups, as quantified by the dimensionless global density parameter, $\eta$ \citep{2013MNRAS.436L..40N}, defined as:
\begin{equation}
        \eta = \frac{\mid \Delta v \mid}{\sigma_{g}} \frac{\Delta r}{R_{200}}
    \label{eq:eta}
    .\end{equation} 
    \\
Here $\Delta v$, $\Delta r$, and $\sigma_{g}$ are the velocity of the galaxy within the group, the projected distance to the group centre, and the velocity dispersion of group members, respectively. $R_{200}$ is the radius where the density of the group is equivalent to 200 times the Universe’s critical density. Galaxies with $\eta$ < 0.1 are considered early accreted by their host group. If they have passed the pericentre of their orbit once 0.1 < $\eta$ < 0.4 is expected, and if they have recently been accreted by the group 0.4 < $\eta$ < 2 is anticipated \citep{2013ApJ...768..118N}. 

We performed a Kolmogorov-Smirnov (K-S) test to quantify the distribution of various stellar parameters between samples of galaxies split by their environments, with the null hypothesis that two samples are drawn from the same distribution. 

The first sample was fixed and corresponds to all galaxies in groups (with all values of $\eta$). Our other samples are: i) galaxies in the field, ii) galaxies in groups with $\eta$ < 0.1, iii) galaxies in groups with 0.1 < $\eta$ < 0.4, and iv) galaxies with $\eta$ > 0.4. Thus, we had four pairs of samples to test against the baseline sample of all galaxies in groups. We tested the stellar mass, t$_{90}$, mean stellar metallicity, and age distributions for all sample pairs mentioned. 

We can reject the null hypothesis (p-values < 0.02) for the metallicity, stellar ages, masses, and t$_{90}$ distributions of galaxies in the field when compared with the baseline sample of all galaxies in groups. This means that the stellar parameter distributions of field and group galaxies are drawn from different distributions. 

In the case when we tested the baseline sample of all galaxies in groups against galaxies in groups with different $\eta$ values, the only case where the null hypothesis is rejected (p-value = 0.004) is for the formation timescale, t$_{90}$, of the sample of galaxies with $\eta$ > 0.4. 

We can conclude that field galaxies are distinct from galaxies in groups, in terms of mean metallicity, ages, stellar masses and formation timescales. However, whether a group galaxy is more dynamically bounded to its group (e.g. lower $\eta$ values) or not, is mostly irrelevant when compared with all the galaxies in groups. The only exception, when we observe two distinct distributions, is when the formation timescales of galaxies with $\eta$ > 0.4 is compared to the overall population of group galaxies in the sample.

In Figure \ref{fig:environment}, we present some stellar population parameters colour-coded by the global density estimator, $\eta$. In the left panel of Figure \ref{fig:environment} we present the stellar mass as a function of how fast 90$\%$ of it was formed; we do not observe any correlation with the environment given that we find galaxies that assemble 90$\%$ of their stellar mass in all covered environments. We do see, looking at the relative fractions, that more massive systems assemble their stellar mass faster, mostly in less than 8 Gyr. Also, the relative fractions for galaxies in less dense environments are more similar to the t$_{90}$ distribution for field galaxies, having an important fraction of galaxies with formation timescales > 8 Gyr. We observe as well, in the right panel of Figure \ref{fig:environment}, that galaxies that form their stellar mass faster (t$_{90}$ < 3 Gyr) tend to have higher metallicities. Thus, high-mass galaxies are formed quickly at high-z, stay quiescent, and have high metallicity regardless of their environment at the epoch of observation. Nevertheless, we do have some massive galaxies at z $\sim$ 0.6 in various environments with lower metallicities and short  t$_{90}$ (< 5 Gyr). These systems are consistent with having high levels of SF and forming most of their stars somewhat later, at z $\sim$2.
In Figure \ref{fig:hist_env}, we present the redshift and stellar mass distributions for field galaxies and galaxies in groups. Overall, we observe a high number of massive galaxies in groups at higher redshifts, compared to galaxies in the field. Thereby, the mass-metallicity (shown in Figure \ref{fig:mass_metal_eta}) relation in our sample is mostly driven by massive galaxies at high-redshift (z $\sim$ 0.7), which are in groups and/or structures.

    \begin{figure}
    \centering
       \includegraphics[width=7.7cm]{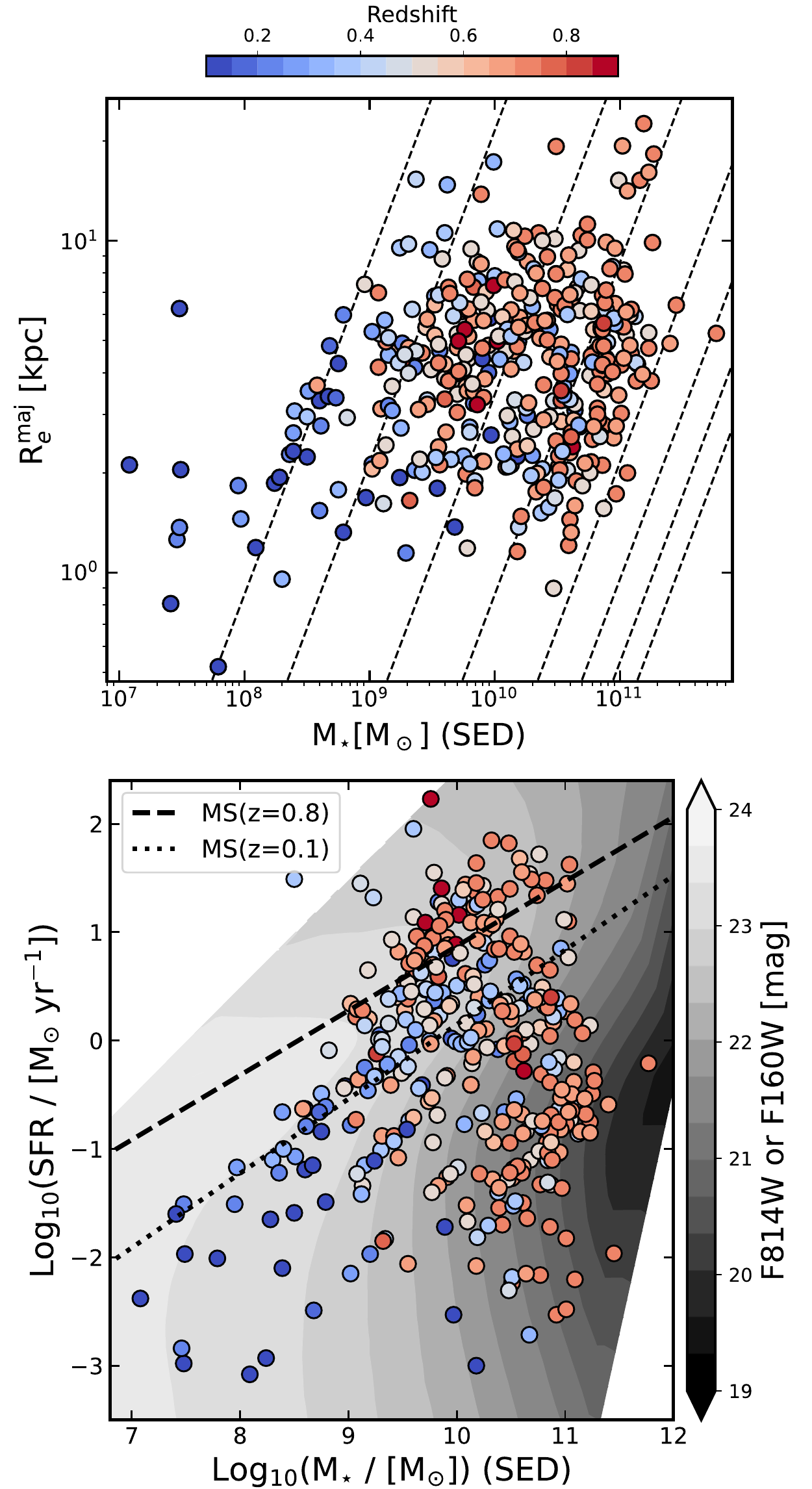}
         \caption{\textit{Top:} Plane of effective radius (R$^{\rm maj}_{\rm e}$) versus stellar mass coloured by SSP-equivalent population parameters measured within one effective radius. Dashed-black lines show constant velocity dispersion of 10, 20, 50, 100, 200, 300, 400, 500 km s$^{-1}$ from left to right, derived by the virial mass estimator M = 5R$^{\rm maj}_{\rm e}$ $\sigma^{2}$/G. \textit{Bottom:} General properties of the galaxy sample and the SFR as a function of the stellar mass (from SED fitting) for the sample galaxies. The dotted and dashed grey lines indicate the main sequence (MS) for an age of the Universe respectively of 12.4 Gyr (z = 0.1) and 6.8 Gyr (z = 0.8), given by Eq. (1) of \cite{2012ApJ...754L..29W}. We extrapolated the MS relations for galaxies with stellar mass M$_{\star}$ $\leq$ 10$^{8.5}$M$_{\odot}$. The grey background is the LOESS smoothed brightness of all objects brighter than 24 mag within the MAGIC and MUSE-Wide footprints.
}
         \label{fig:bias}
    \end{figure}

We analyze the effect of the different stellar masses covered by field galaxies and galaxies in groups on the observed stellar population parameter trends. We used a partial correlation method, implemented by the python library $\mathtt{Pingouin}$ \citep{ppcor}, to study the role of the mass in our results. We tested the mean stellar ages, metallicities, and t$_{90}$ against the environmental parameter $\eta$ after removing the effect of the stellar mass. We found in all the performed tests that the stellar mass is not driving the observed trends with the stellar population parameters.

\subsection{Sample biases}\label{sec:bias}

We would like to compare our results with similar studies using nearby galaxies. However, to do this, we have to analyze how representative of the Local and inter-z Universe our galaxy sample is. 

As the MUSE surveys we used in this work are flux-limited, our sample is limited by the galaxy's integrated S/N. Given the MUSE spatial resolution, the S/N for our sample is influenced not only by the galaxy brightness but also by the angular size. Therefore, as a consequence, the sample has an implicit bias in galaxy mass and size. At low redshifts, we collect galaxies of various sizes and stellar masses, as shown in the top panel of Figure \ref{fig:bias}, where we see that z $\leq$ 0.4 systems range from 10$^{8}$ to 10$^{11.6}$ M$_{\odot}$, and various sizes from compact to large scales. However, as the redshift increases we collect the brightest galaxies. We show this in the bottom panel of Figure \ref{fig:bias}, where the shaded-grey area corresponds to the brightness of $\sim$10000 galaxies in the CANDELS and COSMOS areas (from the 3D-HST program) over the footprint of our sample. The contours represent the underlying distribution of galaxies which were smoothed using the LOESS method. We also observe that on average brighter objects have lower levels of star formation. At high redshifts (z $\geq$ 0.6) our galaxies (shown by the blue-red colour scheme) are bright and large systems, which in terms of stellar mass, corresponds to the most massive ones. 

We are covering the galaxy's mass and SFR ranges that local and higher-redshift surveys target. At redshifts around 0.014 $\leq$ z $\leq$ 0.10, SAMI survey includes galaxies with all morphologies and environment. In terms of stellar masses, the survey ranges from 10$^{10}$M$_{\odot}$ to $\sim$10$^{11.6}$ M$_{\odot}$ and includes both, star-forming and quiescent galaxies. At higher redshifts z $\sim$ 0.8, LEGA-C survey also surveyed star-forming and quiescent galaxies. The stellar mass for LEGA-C systems goes from 10$^{10.4}$M$_{\odot}$ to $\sim$10$^{11.6}$ M$_{\odot}$. In comparison with \cite{2022MNRAS.512.3828B} (their Fig.2), we see that our sample extends both the mass and the SFR to galaxies with mass lower that 10$^{10.7}$M$_{\odot}$ and Log$_{10}$(SFR/[M$_{\odot}$ yr$^{-1}$]) $\leq$ -2, respectively.

We also have low-mass systems (M$_{\star} \leq$ 10$^{9.5}$ M$_{\odot}$), which are galaxies at low redshift, both in the star-forming sequence and also quiescent systems. Nevertheless, our sample lacks galaxies with low masses \citep{2018A&A...619A..27B} and high levels of star formation. Generally, dwarfs star-forming galaxies are populating low-density environment  \citep{1988ApJ...330L..17I, 2000MNRAS.311..307T, 2014MNRAS.445.1694L} and usually present low surface brightness \citep{1993AJ....106.1784C, 1995MNRAS.275....1T}. Therefore, as they are not bright and not massive enough, our sample misses them. 

Therefore, we are sampling starbursting, star-forming, and quenched galaxies equivalent to local and z $\sim$ 0.8 surveys, in the mass range from 10$^{9}$ M$_{\odot}$ to 10$^{12}$ M$_{\odot}$. Nevertheless, we lack star-forming and starbursting galaxies at stellar masses lower than 10$^{9}$ M$_{\odot}$. Accordingly, our sample is representative of galaxies more massive than 10$^{9}$ M$_{\odot}$ for z $\leq$ 0.5 and 10$^{10}$ M$_{\odot}$ for 0.7 $\leq$ z $\leq$ 0.9.


\section{Conclusions}\label{sec:conclusions}

We measured the stellar population parameters and derived the SFH of a sample of 393 galaxies in various MUSE surveys. The sample contains galaxies in the redshift range of 0.1 $\leq$ z $\leq$ 0.9, stellar masses from $\sim$10$^{8}$ M$_{\odot}$ to 10$^{12}$ M$_{\odot}$, and SFRs derived from SED fitting from Log$_{10}$(SFR/[M$_{\odot}$ yr$^{-1}$]) $\approx$ -3 to $\approx$ 2 . The stellar population parameters were computed within an aperture of 1 $R_{e}$ for each galaxy. 

Thanks to our analysis of the mass-size plane, we observed that fixed-mass, compact galaxies are (on average) older and more metal-rich,  and they form their stellar masses faster than their larger counterpart. Most of the galaxies in our sample form a great fraction of the stellar mass when the Universe was young (< 2 Gyr). More massive galaxies have higher levels of star formation at earlier times and reach quiescence faster than lower-mass systems, assembling their total mass earlier, around z $\sim$ 1.

From the spectral fitting, we also derive SFR as a function of cosmic time for the galaxy sample. We modelled the changes in the SFR during the galaxy's life, using a combination of exponential and Gaussian functions. Using this method we identified the main events of star formation in the systems. We found that more than 85$\%$ of the galaxies have more than one SFE, generally described by an exponential and Gaussian distributions. However, the more massive galaxies usually have one main event only, described by exponential decay SFR, followed by a low-level, but continuous SF; or, possibly, mass assembly through accretion. In most cases, the initial SF event happened at high-z (z > 2) and contributed to $\sim$ 50$\%$ of their total stellar mass.

As our sample probes different environments from low-density to massive galaxy structures, we quantified it using the global density $\eta$ estimator. We analysed our results in search of any environmental correlation and found that the main difference can be seen between the nature of a galaxy in the field and a galaxy in a group and/or structure.

As a last step, we considered how representative our
sample is with respect to the Local Universe  on the whole and, in addition, what the possible sample biases could be. As MUSE surveys are flux-limited, we find that more massive galaxies are mostly found at high redshift in our sample. They are also more common in dense environments, as shown in \citet{2024A&A...683A.205E} for the MAGIC survey. In the mass range from 10$^{9}$ M$_{\odot}$ to 10$^{12}$ M$_{\odot}$, we are sampling starbursting, star-forming, and quenched galaxies with a similar approach to other surveys at these redshifts. However, at stellar masses lower than 10$^{9}$ M$_{\odot}$, we lack star-forming and starbursting systems given their dim brightness properties. Therefore, we conclude that our sample is representative of galaxies more massive than 10$^{9}$ M$_{\odot}$ for z $\leq$ 0.5 and 10$^{10}$ M$_{\odot}$ for 0.7 $\leq$ z $\leq$ 0.9. Therefore, the  results presented in this paper should be taken with this caveat included.

\begin{acknowledgements}
      This work was supported by the German
      \emph{Deut\-sche For\-schungs\-ge\-mein\-schaft, DFG\/} project
      number 4548/4-1. CML thanks Wilfried Mercier, Jakob Walcher, Richard Mcdermid, and Michele Cappellari for their help and advice. Many thanks to Max for his emotional support provided during the writing process of this work. 
\end{acknowledgements}

\clearpage
\begin{appendix}
\section{}
    \begin{figure}[H]
    \centering
       \includegraphics[width=8cm]{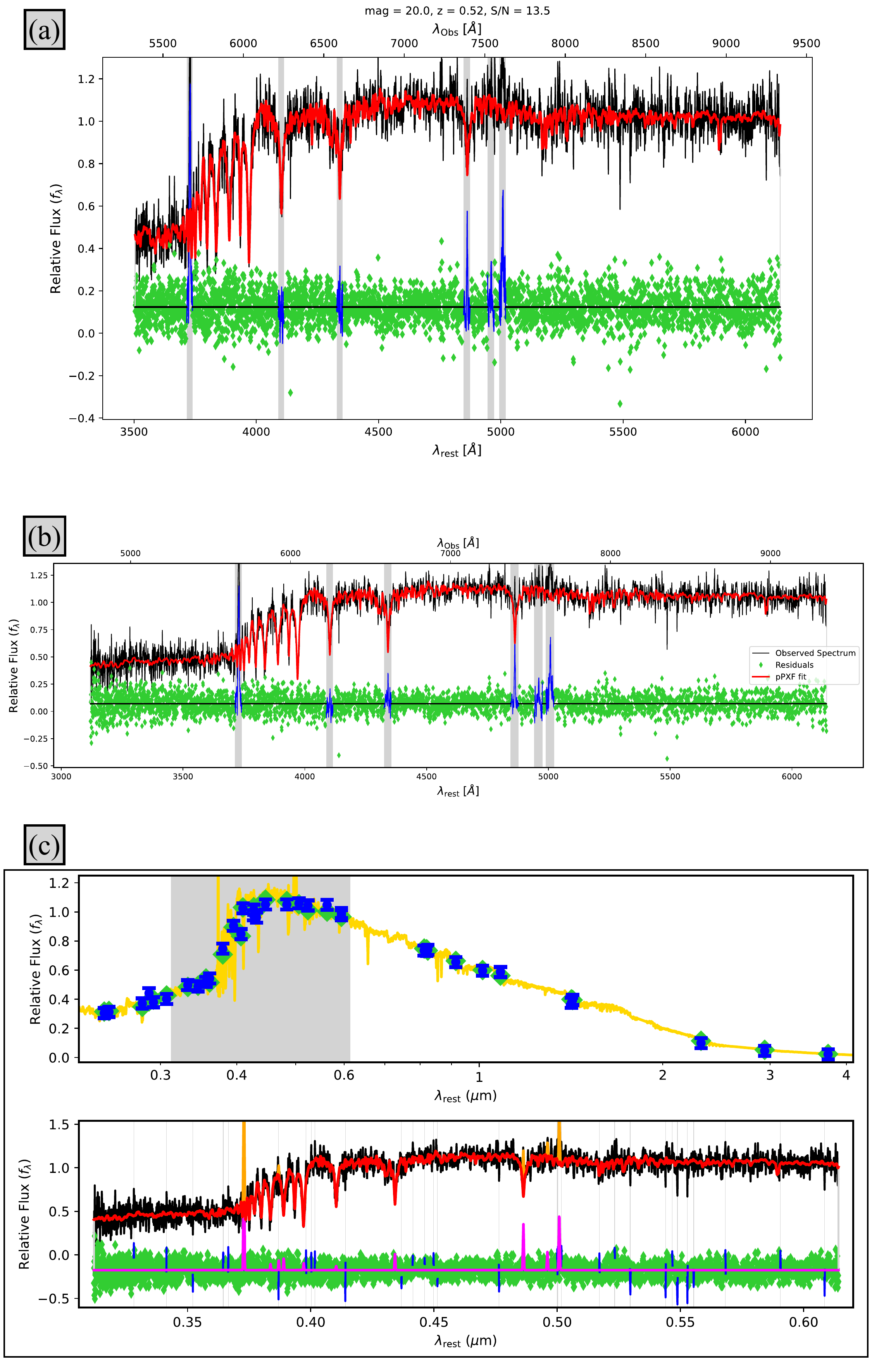}
          \caption{Example of the population analysis for one galaxy in the sample, with index N: 36. \textit{Panel (a):} Stellar fit with Indo-US templates used to extract the global velocity dispersion, estimate S/N and build the galaxy sample. On the top part we show the magnitude in the HST F160W band, the galaxy redshift and the integrated S/N. The black line is the observed spectrum, best stellar fit spectrum is in red, masked emission lines are shown in blue and residuals with green diamonds. \textit{Panel (b):} Observed spectrum in black, pPXF best-fitting total spectrum with the orange line, best stellar fit spectrum in red, and gas emission in magenta. Green diamonds are residuals from the fit and blue lines are masked pixels (shown as grey areas as well). This fit was performed using E-MILES templates with masked emission lines with the purpose of deriving the SFH. \textit{Panel (c):} Best-fitting template using E-MILES templates and including the fit to the emission lines is shown with the golden line, blue-error-bar symbols show photometric measurements and green diamonds indicate the best fit. The grey area corresponds to the wavelength range covered by the spectrum shown in the upper panel. This fit was used to estimate the spectral masses of galaxies, measure gas kinematics and emission-line fluxes.}
         \label{fig:example_36_appendix}
    \end{figure}

    \begin{figure}
    \centering
       \includegraphics[width=8cm]{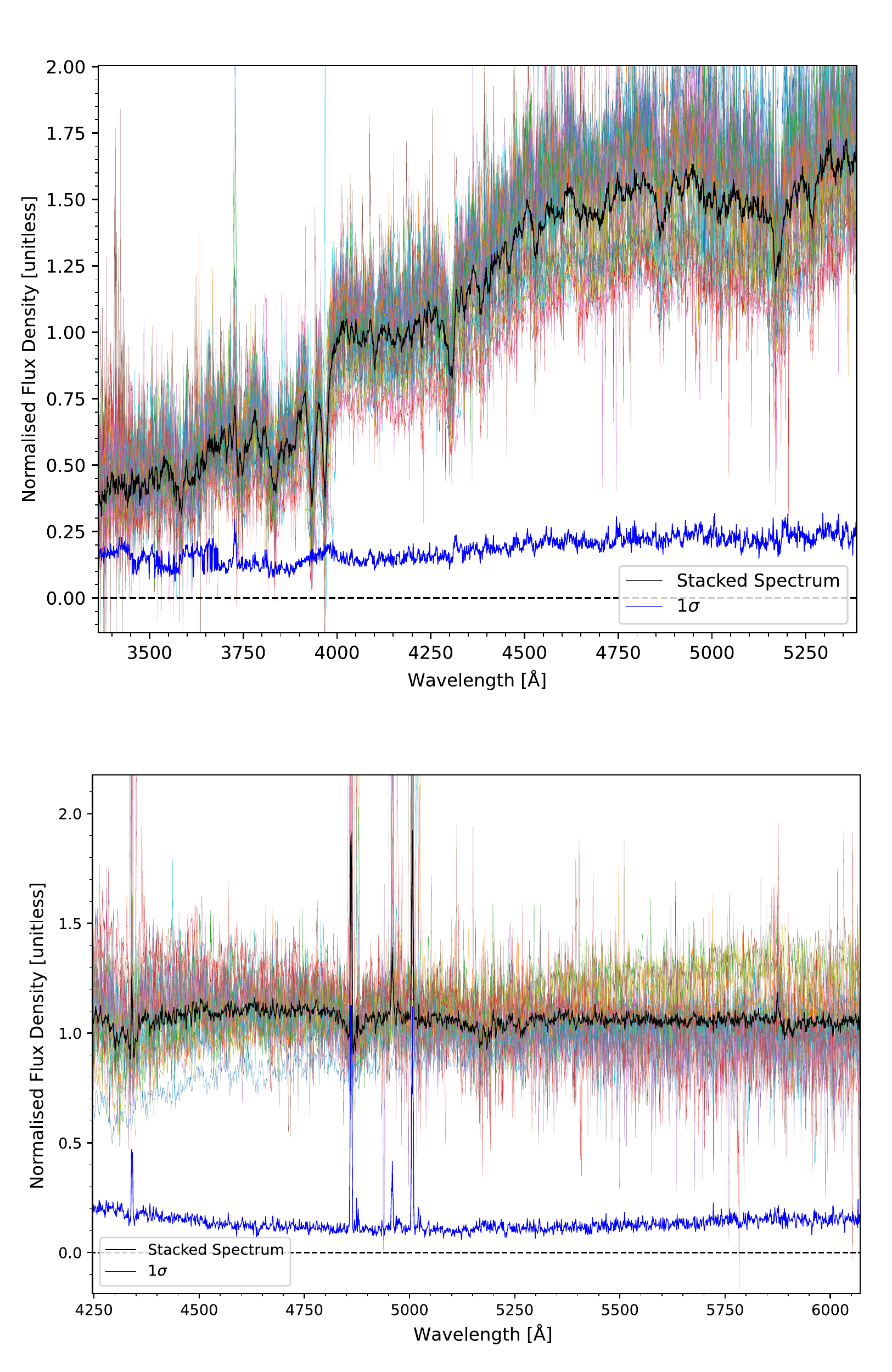}
        \caption{Stacked spectrum (in black line) for galaxies with 1 and 4/5 SFEs, in the upper and bottom panels, respectively. The background spectra correspond to the galaxies used for the stacking ($\sim$35 galaxies for each sample). The 1$\sigma$ lines correspond to the standard deviation of the stacked spectrum. The stacked spectra was retrieved using \texttt{Specstack} \citep{2019ascl.soft04018T}.}
         \label{fig:stacked}
    \end{figure}

    \begin{figure*}
    \centering
       \includegraphics[width=17.2cm]{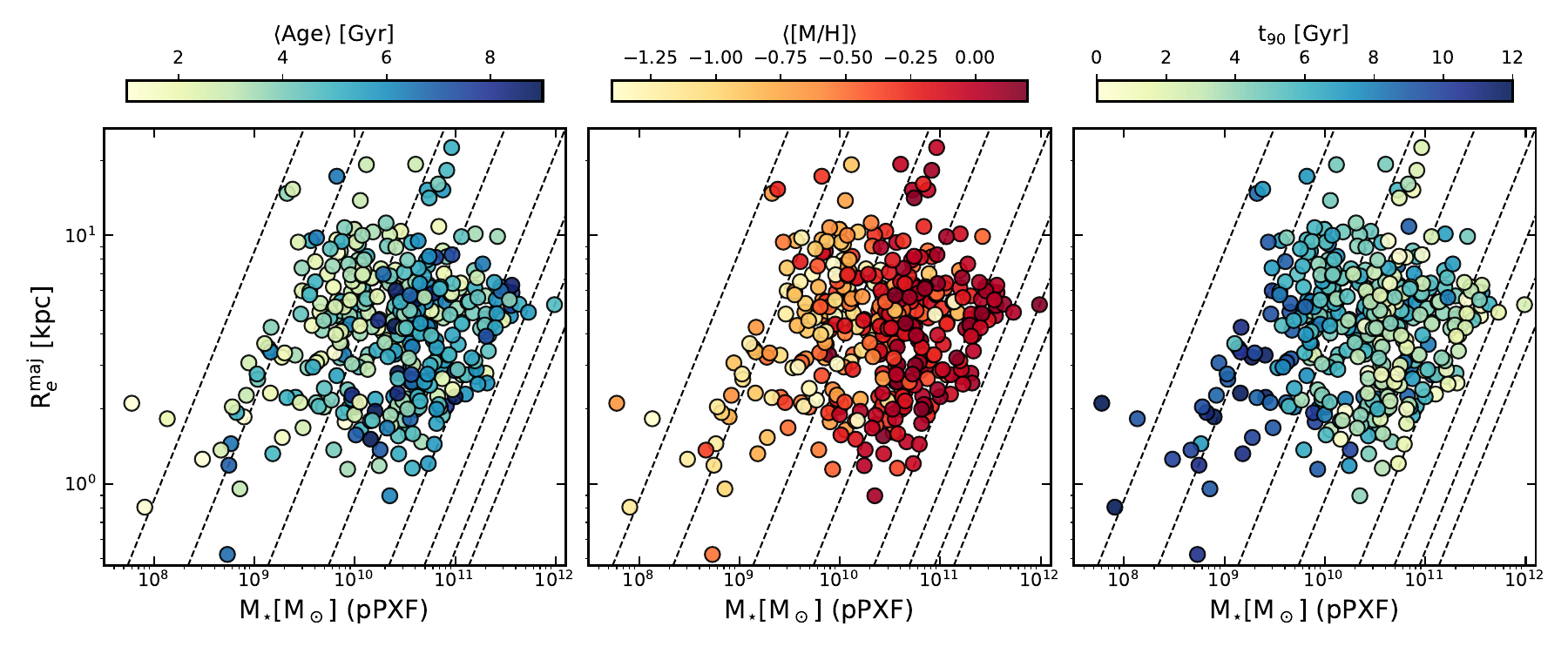}
          \caption{Plane of effective radius (R$^{\rm maj}_{\rm e}$) versus stellar mass coloured by SSP-equivalent population parameters measured within one effective radius. Colours of the left, central, and right panels correspond to the mean stellar age, mean stellar metallicity, and the time taken to form 90$\%$ of the star in the galaxy, respectively. Dashed-black lines show constant velocity dispersion of 10, 20, 50, 100, 200, 300, 400, 500 km s$^{-1}$ from left to right, derived by the virial mass estimator M = 5R$^{\rm maj}_{\rm e}$ $\sigma^{2}$/G. The LOESS smoothed version of this figure is shown in Figure \ref{fig:mass_size}.}
         \label{fig:mass_size_raw}
    \end{figure*}

    \begin{figure}
    \centering
       \includegraphics[width=6cm]{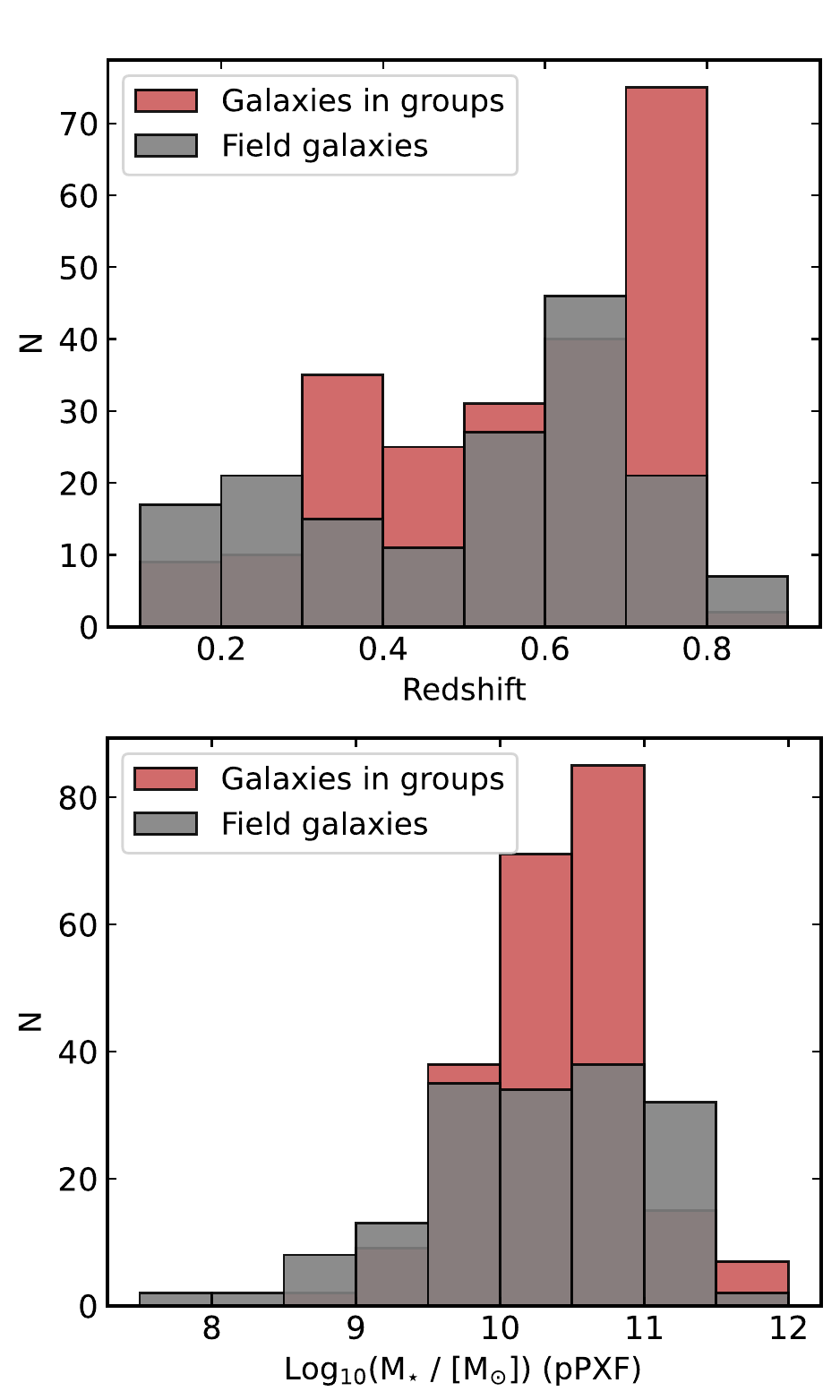}
        \caption{Redshift and stellar mass distributions for galaxies in groups (with different values of the environmental indicator $\eta$) and field galaxies. More details in Section \ref{sec:env}.}
         \label{fig:hist_env}
    \end{figure}
    \begin{figure}
    \centering
       \includegraphics[width=8.6cm]{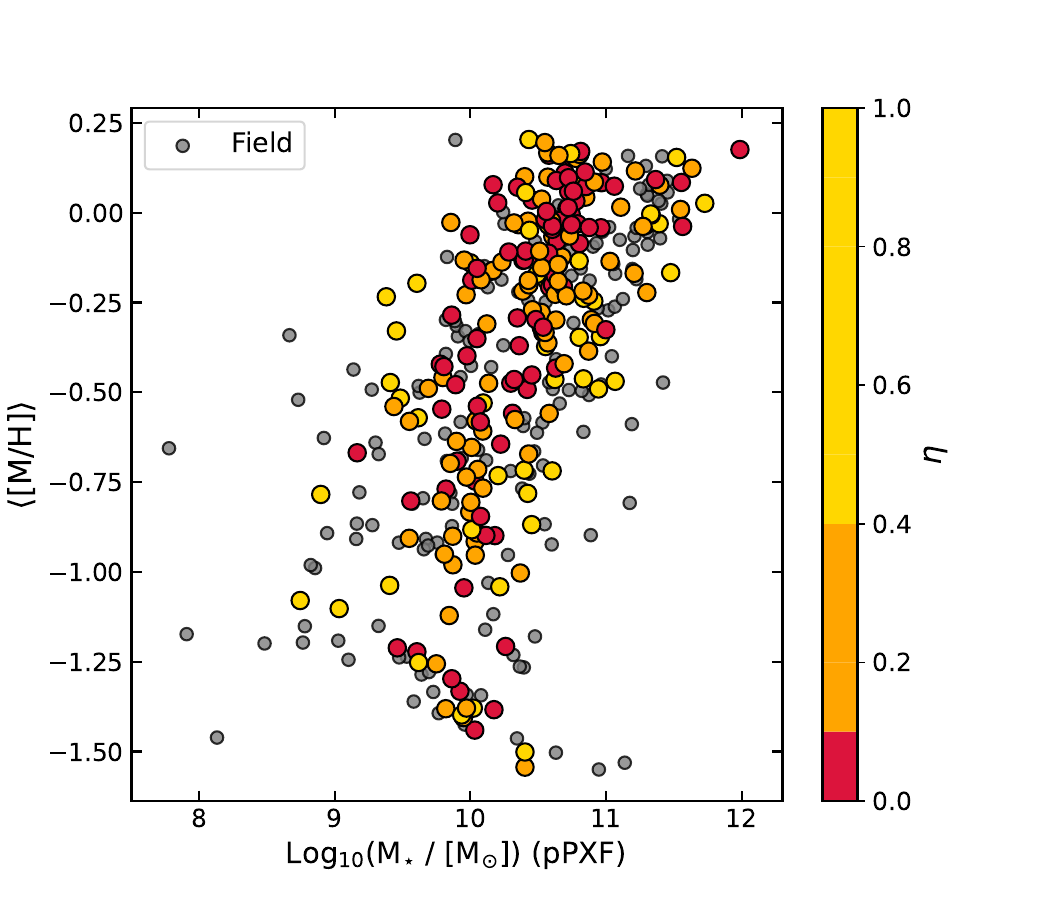}
        \caption{Mass-metallicity relation coloured by the environmental indicator, $\eta$. Grey dots denote field galaxies. More details in Section \ref{sec:env}. From the figure, we can observe that our sample is dominated by massive galaxies at high redshifts (z $\sim$ 0.7) in group environments. Therefore, the mass-metallicity relation is dominated by these type of galaxies.}
         \label{fig:mass_metal_eta}
    \end{figure}

    \begin{figure*}
    \centering
       \includegraphics[width=11.7cm]{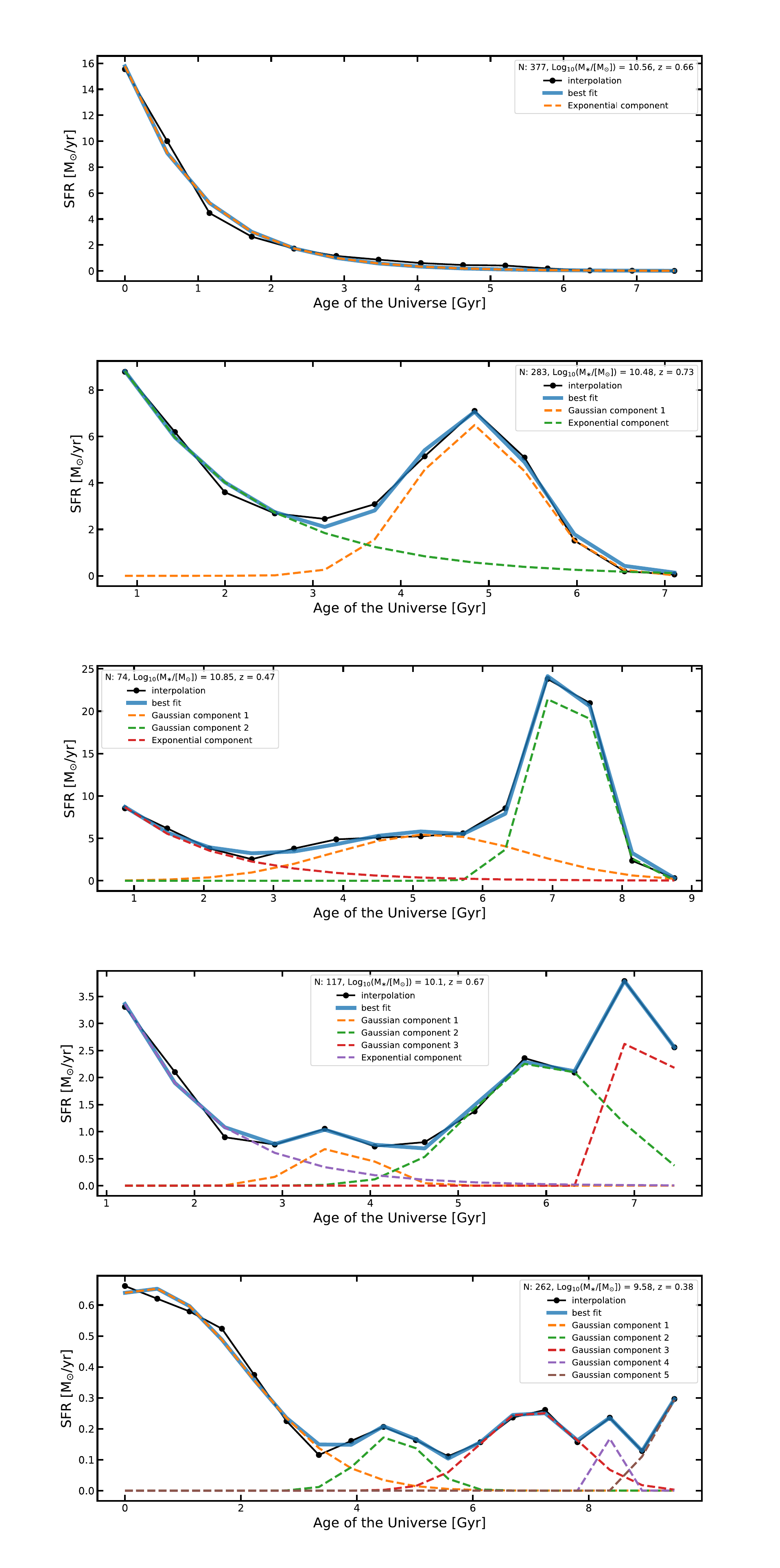}
        \caption{SFH as a function of time for five galaxies in the sample. From top to bottom:\ Galaxies with one, two, three, four, and five SFEs, respectively. In each panel, the black line shows the derived SFR in bins of 0.5 Gyr. Blue-solid line show the best fit of the SFR. Dashed lines show the decomposition of the best fit curve into different SFEs specified in the legend. The galaxy index (N) is also indicated in the legend. More details in Section \ref{sec:sfe}.}
         \label{fig:sfe_example}
    \end{figure*}
\clearpage

\begin{table*}[]
\centering
\begin{tabular}{ |p{3cm}||p{11.5cm}|  }
    \hline
    \multicolumn{2}{|c|}{Galaxy sample properties} \\
    \hline
    \hline
    Parameter& Description\\
    \hline
    \hline
    N  &  Galaxy index\\
    RA  &  RA J2000 degrees of galaxy centre\\
    DEC  &  Dec J2000 degrees of galaxy centre\\
    Z  &  MUSE spectroscopic redshift\\
    SN  &  MUSE S/N ratio of the galaxy measured within the effective radius\\
    Re$\_$maj$\_$kpc  &  Size parameter, R$_{e}^{\rm maj}$, in kpc\\
    lmass$\_$sed  &  Galaxy SFR from SED fitting\\
    lsfr$\_$sed  &  Galaxy stellar mass from SED fitting\\
    lmass$\_$ppxf  &  Galaxy stellar mass from spectral pPXF fitting\\
    lmass$\_$ppxf$\_$err  &  Galaxy stellar mass error from spectral pPXF fitting\\
    l$\_$age  &   Galaxy stellar ages from spectral pPXF fitting\\
    l$\_$age$\_$err  &  Galaxy stellar ages error rate from spectral pPXF fitting\\
    metal   &   Galaxy stellar metallicity from spectral pPXF fitting\\
    metal$\_$err   &  Galaxy stellar metallicity error rate from spectral pPXF fitting\\
    lsfr$\_$OII  &  Galaxy SFR from [OII] doublet emission\\
    t$\_$50   &  Time taken to assemble 50$\%$ of the stellar mass\\
    t$\_$90   &  Time taken to assemble 90$\%$ of the stellar mass\\
    eta   &  Environmental global density parameter\\
    N$\_$SFE   & Number of SFEs\\
    SFE   & Types of SFEs\\ \hline
\end{tabular}
\caption{Galaxy sample properties. The full Table is available at the CDS.}
\label{tab:sample_table} 
\end{table*}

\end{appendix}
\end{document}